\documentclass[12pt,preprint]{aastex}
\usepackage[dvips]{color}
\usepackage{longtable,lscape}
\usepackage[usenames,dvipsnames]{xcolor}
\usepackage{amsmath}    

\begin{document}

\newpage
\newcommand{\cd}     {CD~$-38^{\circ}\,245$}
\newcommand{\cdsean} {CD~$-24^{\circ}\,17504$}
\newcommand{\bdbm}   {BD~$+18^{\circ}\,5550$}
\newcommand{\hen}    {HE~0107--5240}
\newcommand{\hea}    {HE~1327--2326}
\newcommand{\hej}    {HE~0557--4840}
\newcommand{\hez}    {HE~0057--5959}
\newcommand{\hef}    {HE~1506--0113}
\newcommand{\het}    {HE~2139--5432}
\newcommand{\kms}   {\rm km~s$^{-1}$} 
\newcommand{\teff}  {$T_{\rm eff}$} 
\newcommand{\logg}  {$\log g$} 
\newcommand{\loggf} {log~$gf$} 
\newcommand{\feh}   {[Fe/H]}
\newcommand{\alphafe} {[$\alpha$/Fe]}
\newcommand{\uf}    {ultra-faint}
\newcommand{\mv}    {$M_{V}$}
\newcommand{\ebv}    {$E(B-V)$}
\newcommand{\oodlte} {$_{\rm 1D,LTE}$}
\newcommand{\odlte}  {1D,LTE}
\newcommand{\odnlte}  {1D,NLTE}
\newcommand{\tdlte}  {3D,LTE}
\newcommand{\tdnlte}  {3D,NLTE}
\newcommand{\msun} {\rm M$_\odot$}
\newcommand{\ltsima} {$\; \buildrel < \over \sim \;$}
\newcommand{\simlt} {\lower.5ex\hbox{\ltsima}}
\newcommand{\gtsima} {$\; \buildrel > \over \sim \;$}
\newcommand{\simgt} {\lower.5ex\hbox{\gtsima}}

\title{\bf The Most Metal-Poor Stars. V. The CEMP-no Stars in 3D {\it and} Non-LTE}

\author{John E. Norris\altaffilmark{1} and David Yong\altaffilmark{1,2}}

\altaffiltext{1}{Research School of Astronomy and Astrophysics, The
  Australian National University, Canberra, ACT 2611, Australia;
  jen@mso.anu.edu.au, yong@mso.anu.edu.au}

\altaffiltext{2}{ARC Centre of Excellence for All Sky Astrophysics in 3
Dimensions (ASTRO 3D), Australia}

\begin{abstract}

We explore the nature of carbon-rich ([C/Fe]$_{\rm 1D,LTE}$ $>$ +0.7),
metal-poor ([Fe/H$_{\rm 1D,LTE}$] $<$ --2.0) stars in the light of
post 1D,LTE literature analyses, which provide 3D--1D and NLTE--LTE
corrections for iron, and 3D--1D corrections for carbon (from the CH
G-band, the only indicator at lowest [Fe/H]).  High-excitation C~I
lines are used to constrain 3D,NLTE corrections of G-band analyses.
Corrections to the 1D,LTE compilations of Yoon et al. and Yong et
al. yield 3D,LTE and 3D,NLTE Fe and C abundances.  The number of
CEMP-no stars in the Yoon et al. compilation (plus eight others)
decreases from 130 (1D,LTE) to 68 (3D,LTE) and 35 (3D,NLTE).  For
stars with --4.5 $<$ [Fe/H] $<$ --3.0 in the compilation of Yong et
al., the corresponding CEMP-no fractions change from 0.30 to 0.15 and
0.12, respectively.

We present a toy model of the coalescence of pre-stellar clouds of the
two populations that followed chemical enrichment by the first
zero-heavy-element stars: the C-rich, hyper-metal-poor and the
C-normal, very-metal-poor populations.  The model provides a
reasonable first-order explanation of the distribution of the 1D,LTE
abundances of CEMP-no stars in the $A$(C) and [C/Fe] vs. [Fe/H]
planes, in the range --4.0 $<$ [Fe/H] $<$ --2.0.

The Yoon et al. CEMP Group I contains a subset of 19 CEMP-no stars
(14\% of the group), 4/9 of which are binary, and which have large
[Sr/Ba]$_{\rm 1D,LTE}$ values.  The data support the conjectures of
\citet{hansent16, hansen19} and \citet{arentsen18} that these stars
may have experienced enrichment from AGB stars and/or ``spinstars''.

\end{abstract}


{\noindent \it {The predicted 3D corrections for molecular lines would also alleviate the extraordinarily large C abundances found in some very metal-poor stars. ... Similarly, the fraction of strongly C-enhanced ([C/Fe] {\gtsima} +1.0) metal-poor stars is likely substantially less than the claimed $>$20\% for [Fe/H] {\ltsima} --2.0. ...}\vspace{13mm}} 
\hspace{1.5cm}-- M. Asplund (2005)

\section{Introduction} \label{sec:intro}

The CEMP-no sub-population of carbon-enhanced metal-poor (CEMP) stars,
with [C/Fe] $>$ +0.7 but no enhancement of heavy neutron-capture
elements, arguably contains the most chemically primitive objects
currently known.  Indeed, among the 12 of the 14 metal-poor stars that
have [Fe/H] $<$ --4.5 and [C/Fe] $>$ +1.0, 11 have [C/Fe] $>$ +3.0,
while three have values +1.6, $<$+0.9, and $<$+1.8 dex (see
Tables~\ref{tab:tab1} and ~\ref{tab:tab6} for details).  At least 12
of the 14 belong to the CEMP-no group.  It has been argued that the
latter objects formed within a few hundred million years after the Big
Bang, and probably did so before the carbon-normal stars ([C/Fe] $<$
+0.7 and [Fe/H] {\gtsima} --4.5). Table~\ref{tab:tab1} presents a list
of some 23 major milestones on the nature of these objects and the
role they play in our understanding of the early Universe.
Figure~\ref{fig:fig1} shows the dependence of the carbon abundance,
$A$(C)$_{\rm 1D,LTE}$, and relative carbon abundance, [C/Fe]$_{\rm
  1D,LTE}$, as a function of [Fe/H]$_{\rm 1D,LTE}$ for these CEMP-no
stars, together with their carbon distribution as a function of
[Fe/H]$_{\rm 1D,LTE}$. For definitions, and recent reviews and
introductions to the extensive literature on CEMP-no stars, we refer
the reader to \citet{beers05}, \citet{norris13}, \citet{fn15},
\citet{hansent16}, \citet{yoon16}, and \citet{matsuno17}.

The data in Figure~\ref{fig:fig1} are taken from the literature-based
compilation of \citet{yoon16} in which the carbon abundances of the
CEMP-no stars were determined by analysis of the G-band of the CH
molecule at $\lambda$4300~{\AA}\footnote{We note for future reference
  that for some of the CEMP-s stars in the Yoon et al. sample the
  carbon abundances are based on the C$_{2}$ molecule rather than on
  CH.}, [Fe/H] is based essentially on Fe~I lines, and model
atmosphere techniques together with the almost universally adopted one
dimensional (1D) and Local Thermodynamic Equilibrium (LTE) assumptions
(hereafter {\odlte}) were used.  While this technique has proved very
useful in the past, in particular in differential analyses based on
atomic species, it is not necessarily the case for molecular features.
As emphasized by Asplund and coworkers (see \citealt{asplund05}),
errors of order $\Delta$$A$(C) = --1 dex might be expected in
extremely metal-poor stars as a result of the 1D assumption; and as
highlighted in the above introductory quotation, one should be alive
to the possibility that some of the apparent characteristics of the
CEMP-no stars might result from the {\odlte} assumptions made in the
analysis, rather than the potentially more realistic 3D and non-LTE
(hereafter {\tdnlte}) ones.  More generally, while 1D,LTE is currently
a more precise formalism than 3D,NLTE (which is a much more
challenging endeavor) it is the latter that will result in more
accurate results.

The aim of the present work is to use literature-based corrections
determined by adopting the assumptions of {\tdnlte} to correct carbon
and iron abundances based on those of {\odlte}.  The outline of the
paper is as follows. In Section~\ref{sec:semantics} we address
semantics of carbon richness relevant to the present work.
Section~\ref{sec:corrections} summarizes results from the literature
for 3D--1D,LTE corrections for the analysis of both the G-band and
Fe~I lines and 3D--1D,NLTE corrections for Fe~I lines.  In order to
address the problem that determination of NLTE corrections is not
currently possible for the CH molecule, we also use abundances from
infrared high-excitation C~I lines to constrain the CH NLTE
corrections in the range --3.3 $<$ [Fe/H] $<$ --2.0.  In
Section~\ref{sec:revisions} we use these corrections to update the
1D,LTE Fe and C abundances of \citet{yoon16}, \citet{yong13a}, and a
few more recent values, to place them within the 3D,LTE and 3D,NLTE
frameworks.  As foreshadowed by \citet{asplund05} the changes are
large, and in Section~\ref{sec:yong13b}, following \citet{yong13b}, we
address their effect on the Metallicity Distribution Function and the
fraction of CEMP (principally CEMP-no) stars in the range [Fe/H] $<$
--3.0.  For completeness, Section~\ref{sec:uncertainties} discusses
some uncertainties of the present work, while Section~\ref{sec:naba}
addresses 1D,LTE abundances of the light elements Na, Mg, Al, and Ca,
together with the heavy neutron-capture elements Sr and Ba, and their
implications for the nature of the CEMP-no stars.  In
Section~\ref{sec:toymod} we present a toy model that seeks to explain
the CEMP-no stars in the abundance range --4.0 $<$ [Fe/H] $<$ --2.0 in
terms of the coalescence of gas clouds of C-rich material of the
second generation ([Fe/H] $<$ --5.0, [C/Fe] {\simgt} +1.0), and those
of the C-normal stars of the canonical halo population ([Fe/H] $>$
--4.0, [C/Fe] = 0.0).  Section~\ref{sec:summary} summarizes our
results.

\section{\it The Semantics of Carbon Richness} \label{sec:semantics}

Just what does one mean by carbon richness?  In almost all discussions
based on the analysis of the G-band strength in the spectra of
metal-poor stars with [Fe/H] {\simlt} --3.0, the framework is based on
1D,LTE assumptions, and a star is C-rich if it has an abundance
[C/Fe]$_{\rm 1D,LTE}$ $>$ +0.7 (following \citealp{beers05} and
\citealp{aoki07}).  If, however, 1D,LTE-based results were to
overestimate carbon abundances, by, say, 0.7 dex, stars ``observed''
at this limit would in reality have the solar relative carbon
abundance.  If one is interested in abundances relative to the sun,
it thus follows there is a problem in defining an abundance limit
based on a formalism that has systematic errors which are a function
of chemical abundance.  It would be better to choose an independent
limit relative to the solar abundance that would be useful when
seeking to compare stellar overabundances with, for example,
overabundances that might be observed in other fields, such as gaseous
nebulae and far-field cosmology, and also theoretical models of
stellar, galactic, and cosmological formation and evolution.

Insofar as we shall be discussing carbon and iron abundances
determined using different assumptions we adopt the following
definitions.  As noted above, [Fe/H]$_{\rm 1D,LTE}$ and [C/Fe]$_{\rm
  1D,LTE}$ refer to values determined assuming 1D,LTE.  [Fe/H]$_{\rm
  3D,LTE}$ and [C/Fe]$_{\rm 3D,LTE}$ assume 3D,LTE, and [Fe/H]$_{\rm
  3D,NLTE}$ and [C/Fe]$_{\rm 3D,NLTE}$ adopt 3D,NLTE.  [Fe/H] and
[C/Fe] are used generically. Finally, we adopt a generic carbon
overabundance limit for all of these cases that somewhat arbitrarily
defines carbon richness as [C/Fe] $>$ +0.7 as the independent limit.

\section{\it 3D and Non-LTE Corrections} \label{sec:corrections}

In order to convert the available 1D,LTE carbon and iron abundances of
very metal-poor stars to include 3D and NLTE effects, we seek
corrections of the form $\Delta$$A$(X)$_{\rm 3D,NLTE - 1D,LTE}$ =
$A$(X)$_{\rm 3D,NLTE}$ -- $A$(X)$_{\rm 1D,LTE}$ for analyses of the CH
G-band (X = C) and Fe lines (X = Fe){\footnote{$A$(X) is defined in
    terms of $\epsilon$(X), the abundance of element X, and the
    numbers N$_{\rm X}$ and N$_{\rm H}$ of atoms X and H: $A$(X) =
    log$\epsilon$(X) = log(N$_{\rm X}$/N$_{\rm H}$) + 12.0.  Also, by
    definition, [X/H] = log(N$_{\rm X}$/N$_{\rm H}$)$_\star$ --
    log(N$_{\rm X}$/N$_{\rm H}$)$_\odot$. Here we adopt [Fe/H] =
    $A$(Fe) -- 7.50 and [C/H] = $A$(C) -- 8.43, following
    \citet{asplund05}.}.  The enormous computational challenge to this
  requirement is highlighted by the very small number of relevant
  papers available in the literature.  Further, in most cases, one
  finds partial solutions involving changes between only 3D and 1D,
  assuming LTE ($\Delta$$A$(X)$_{\rm(3D-1D),LTE}$ = $A$(X)$_{\rm
    3D,LTE}$ -- $A$(X)$_{\rm 1D,LTE}$), or between only NLTE and LTE,
  assuming 1D ($\Delta$$A$(X)$_{\rm 1D,(NLTE-LTE)}$ = $A$(X)$_{\rm
    1D,NLTE}$ -- $A$(X)$_{\rm 1D,LTE}$).  As noted above, in the case
  of carbon abundances determined from analysis of the CH G-band, NLTE
  corrections are not currently possible. To cite \citet{gallagher16}
  ``computing full 3D ... NLTE ... departures for molecular data
  ... has not been attempted in great detail because of the
  complexities involved''.

With this in mind we first discuss what is currently possible in the
analysis of the G-band, together with results for Fe~I lines.
Following this, we consider the analysis of near-infrared
high-excitation C~I lines in metal-poor stars, in order to place
constraints on the role of NLTE in determining $A$(C)$_{\rm 3D,NLTE}$
values based on analysis of the G-band.

\subsection{\it 3D and NLTE corrections for the G-band and Fe~I lines} \label{sec:ch_subsec1}

Literature information that we shall use is presented in
Table~\ref{tab:tab2}, where Columns (1) -- (3) contain the star or
model name, {\teff}, and {\logg}, respectively, while Columns (4) --
(9) present [Fe/H]$_{\rm 1D,LTE}$, [Fe/H]$_{\rm 1D,NLTE}$,
[Fe/H]$_{\rm 3D,LTE}$, [Fe/H]$_{\rm 3D,NLTE}$, $A$(C)$_{\rm 1D,LTE}$,
and $A$(C)$_{\rm 3D,LTE}$.  The final column contains source
information.

There are nine cases for which CH corrections are available, from
the work of \citet{collet06, collet07, collet18}, \citet{frebel08},
\citet{spite13}, and \citet{gallagher16}.  Six of the nine cases are
based on analysis of stars, while three are determined entirely from
model atmosphere comparisons.  The data are also plotted in
Figure~\ref{fig:fig2}, where the upper panel (a) presents
$\Delta$$A$(C)$_{\rm(3D-1D),LTE}$ = $A$(C)$_{\rm 3D,LTE}$ --
$A$(C)$_{\rm 1D,LTE}$ versus [Fe/H]$_{\rm 1D,LTE}$. Red and green
symbols refer to dwarfs and giants (defined here to have {\logg}
larger or smaller than 3.35), respectively. The full line in the
figure represents the linear least-squares best fit to the data, which
is given by: $\Delta$$A$(C)$_{\rm(3D-1D),LTE}$ = 0.087 + 0.170~[Fe/H]
(9 points, with RMS = 0.24).

Further literature data are available that provide 1D,LTE
corrections of Fe~I.  Results from the work of
\citet{amarsi16}, \citet{collet06, collet07, collet18},
\citet{ezzeddine17}, and \citet{frebel08} are presented in Columns (4)
-- (7) of Table~\ref{tab:tab2}, where there are 22 stars, all having
1D,NLTE-LTE corrections; seven have 3D,LTE information; and four
have 3D,NLTE data.

The 3D--1D,LTE and 1D,NLTE-LTE corrections for Fe are plotted in
Figure~\ref{fig:fig2}, panel (b) as a function of [Fe/H]$_{\rm
  1D,LTE}$.  The panel also presents linear and quadratic
least-squares lines of best fit, the equations for which are:
$\Delta$[Fe/H]$_{\rm 1D,(NLTE-LTE)}$ = [Fe/H]$_{\rm 1D,NLTE}$ --
[Fe/H]$_{\rm 1D,LTE}$ = 0.013 -- 0.011~[Fe/H]$_{\rm 1D,LTE}$ +
0.019~[Fe/H]$_{\rm 1D,LTE}$$^{2}$ (22 points, with RMS = 0.09) and
$\Delta$[Fe/H]$_{\rm(3D-1D),LTE}$ = [Fe/H]$_{\rm 3D,LTE}$ --
[Fe/H]$_{\rm 1D,LTE}$ = 0.061 + 0.053~[Fe/H]$_{\rm 1D,LTE}$ (7 points,
with RMS = 0.02).

We conclude this section with two comments.  First, as the reader can
confirm from inspection of the middle panel (b) of
Figure~\ref{fig:fig2}, the (3D--1D),LTE corrections are in the
opposite sense to those for the 1D,(NLTE-LTE) case.  Second,
inspection of panels (a) and (b) of Figure~\ref{fig:fig2} reveals that
there appears to be no significant difference between the
distributions of the dwarf and giant stars.  In what follows, we shall
assume that this is the case.\vspace{-4mm}

\subsection{\it [Fe/H] 3D,NLTE Corrections} \label{sec:fe_subsec2}

The previous section presents [Fe/H]$_{\rm 3D,LTE}$ and [Fe/H]$_{\rm
  1D,NLTE}$ corrections (relative to [Fe/H]$_{\rm 1D,LTE}$)
abundances.  What we would really like, however, are [Fe/H]$_{\rm
  3D,NLTE}$ corrections.  The very limited available data in
Table~\ref{tab:tab2} are from \citet{amarsi16} and are shown in the
bottom panel (c) of Figure~\ref{fig:fig2}.  The grey symbols are the
[Fe/H] (3D--1D),LTE and 1D,(NLTE-LTE) corrections from the middle
panel of the figure, while the square symbols above them are the
[Fe/H]$_{\rm 3D,NLTE}$ -- [Fe/H]$_{\rm 1D,LTE}$ corrections.  A very
significant result of the bottom panel (c) is that while the
[Fe/H]$_{\rm 1D,NLTE}$ -- [Fe/H]$_{\rm 1D,LTE}$ corrections in (b) are
positive and the [Fe/H]$_{\rm 3D,LTE}$ -- [Fe/H]$_{\rm 1D,LTE}$
values, also in (b), are negative, the [Fe/H]$_{\rm 3D,NLTE}$ --
[Fe/H]$_{\rm 1D,LTE}$ corrections in (c) are positive and larger than
[Fe/H]$_{\rm 1D,NLTE}$ -- [Fe/H]$_{\rm 1D,LTE}$ (from (b)), by 0.08 --
0.15 dex (with a mean value 0.12).  This suggests that when 3D and
NLTE effects are treated in a self-consistent manner the NLTE
corrections dominate.  In the absence of other information, in the
following we shall assume that [Fe/H]$_{\rm 3D,NLTE}$ -- [Fe/H]$_{\rm
  1D,NLTE}$ = 0.12\footnote{A similar effect was reported by
  \citet[Table~3]{nordlander17} in their 3D,NLTE analysis of
  SMSS~0313--6708 (the most iron-poor star currently known, with
  [Fe/H]$_{\rm 3D,NLTE}$ $<$ --6.5), in which they report that the
  [Fe/H]$_{\rm 3D,NLTE}$ -- [Fe/H]$_{\rm 1D,LTE}$ correction is 0.20
  dex.}.

\subsection{\it High-excitation C~I lines and 3D,NLTE corrections for CH} \label{sec:ci}

As emphasized above, {\tdnlte} carbon abundances based on the analysis
of the G-band are currently unavailable due to the intractability of
the CH molecule to NLTE analysis.  The near-infrared, high-excitation C~I
lines, however, are not affected by this problem.  With excitation
potentials $\sim$~7 eV, these lines form deep in the stellar
atmosphere, well below the outer layers where the 3D effects are
significant.  We now use literature C~I abundance analyses to obtain
estimates of 3D,NLTE corrections for CH-based values.  By comparing
the results for stars for which carbon abundances have been obtained
from analyses of both the CH G-band and near-infrared C~I lines, we
then estimate the sense and size of the {\tdnlte} corrections for the
CH-based carbon abundances discussed in the previous section.

\subsubsection{\it Carbon abundances from the near-infrared C~I lines}

\citet{fabbian09} present $A$(C)$_{\rm 1D,NLTE}$ abundances for 43
metal-poor dwarfs and subgiants in the abundance range --3.2 $<$
[Fe/H] $<$ --1.3, based on analysis of the high-excitation C~I 9094.8
and 9111.8~{\AA} lines (EP = 7.49 eV).  They also provide atmospheric
parameters {\teff}, {\logg}, and [Fe/H]$_{\rm 1D,LTE}$, together with
[C/H]$_{\rm 1D,LTE}$, and [C/H]$_{\rm 1D,NLTE}$ for two values of the
Drawin scaling factor S$_{\rm H}$ (= 0.0 and 1.0).  In what follows we
shall adopt the average of these two values of [C/H] $_{\rm 1D,NLTE}$.
\citet{fabbian09} also noted that the high-excitation potential of the
C~I lines would very likely lead to only small (3D--1D),NLTE
corrections, given that these lines are formed sufficiently deep in
the atmospheres of the stars to be insensitive to the 3D effects,
which are significant principally in the outermost layers.  This
expectation is supported by the work of \citet{dobrov13} from their
comprehensive analysis of (3D--1D),LTE corrections for a large number
of atomic species as a function of excitation potential, among other
parameters.  In particular their Figure 4 shows that for neutral
carbon lines having EP = 6eV, $\Delta$(3D--1D),LTE = 0.04 dex.  That
is, effectively, $A$(CI)$_{\rm 1D,NLTE}$ = $A$(CI)$_{\rm
  3D,NLTE}$.\footnote{Towards the completion of the present work,
  \citet{amarsi19} presented a 3D,NLTE re-analysis of the
  \citet{fabbian09} dataset. Comparison of the 1D,NLTE carbon
  abundances of these two works (their Figures 1 and 5, respectively)
  show good agreement to within $\sim$~0.1 dex, while the
  \citet{amarsi19} 1D,NLTE and 3D,NLTE values differ by {\simlt}~0.1
  dex.} (For convenience, in this sub-section we shall refer to
abundances based on C~I lines as $A$(CI) and those on the CH features
as $A$(CH).)

To proceed further we also require CH-based carbon abundances for
these stars.  For 23 of the Fabbian et al. sample we obtained
high-resolution, high signal-to-noise spectra from astronomical
archives.  Details of this sub-sample are presented in
Table~\ref{tab:tab3}.  Columns (1) -- (4) contain the star name,
{\teff}, {\logg}, and [Fe/H] from \citet{fabbian09}, while columns (5)
-- (6) present their values of $A$(CI)$_{\rm 1D,LTE}$ and $A$(CI)$_{\rm
  1D,NLTE}$.  Columns (10) -- (11) of the table contain the $S/N$ of the
spectra and the archives from which the CH data were obtained.

To obtain $A$(CH) we proceeded as follows.  For each star we co-added
multiple spectra as available, followed by continuum normalization.
Using the atmospheric parameters of \citet{fabbian09},
model-atmospheric spectra of each star were computed for several
carbon abundances over the range 4305 -- 4330~{\AA}. We refer the
reader to \citet{yong13a} for details of the technique. In brief, we
used the code MOOG \citep{sneden73}, as modified by \citet{sobeck11},
together with the model atmospheres of \citet{castelli03}.  In the
linelist, the data for CH lines were provided by B. Plez et al. (2009,
private communication; see \citealp{masseron14}). Other pertinent data
are: we adopted microturbulence = 1.5 {\kms} and [O/Fe] = 0.40, and
note that the results are insensitive to the latter, given the
relatively high effective temperatures of these dwarfs.  The resulting
abundances are presented in Table~\ref{tab:tab3}, where columns (7) --
(9) contain $A$(CH)$_{\rm 1D,LTE}$, $A$(CH)$_{\rm 3D,LTE}$, and
$A$(CH)$_{\rm 3D,LTE}$ -- $A$(CH)$_{\rm 1D,LTE}$, respectively
($A$(CH)$_{\rm 3D,LTE}$ was computed using the (3D--1D),LTE
corrections presented in Section~\ref{sec:ch_subsec1} above).  For
comparison purposes, we also present in Table~\ref{tab:tab3} the CH
based literature abundances for {\cdsean} from \citet{jacobson15}, and
G64-12 and G64-37 from \citet{placco16}.  We note that the mean
difference between our results and those of \citet{jacobson15} and
\citet{placco16} is $\langle$$\Delta$$A$(CH)$_{\rm 1D,LTE}$$\rangle$ =
0.06.

\subsubsection{\it Estimating the 3D,NLTE corrections appropriate for the CH G-band}

We now estimate the sense and size of 3D,NLTE corrections appropriate for
the CH G-band analysis.  In Figure~\ref{fig:fig3} the top two panels
(a and b) present $A$(CI)$_{\rm 1D,LTE}$ vs. $A$(CH)$_{\rm 1D,LTE}$
and $A$(CI)$_{\rm 3D,LTE}$ vs. $A$(CH)$_{\rm 3D,LTE}$, respectively,
for heuristic purposes, to give the reader a feeling for the changes
brought about by 3D and NLTE effects. The bottom panel (c) of the
figure shows $\Delta$$A$(C) = $A$(CI)$_{\rm 3D,NLTE}$ -- $A$(CH)$_{\rm
  3D,LTE}$ as a function of [Fe/H]$_{\rm 1D,LTE}$. If the C~I and CH
estimates of carbon abundances accurately and self-consistently
include all 3D and NLTE effects, $\Delta$$A$(C) should be zero.  Given
(as discussed above) that $A$(CI)$_{\rm 1D,NLTE}$ = $A$(CI)$_{\rm
  3D,NLTE}$, any departure of $\Delta$$A$(C) from zero represents an
estimate of the CH 3D,NLTE corrections needed for these stars.  The
negative values of $\Delta$$A$(C) for G64-12 and G64-37 in
Figure~\ref{fig:fig3} indicate that their $A$(CH)$_{\rm 3D,LTE}$
values are larger than $A$(CH)$_{\rm 3D,NLTE}$.  That is, a further
negative correction is required to produce more accurate $A$(CH)$_{\rm
  3D,NLTE}$ values.  The full red line in Figure~\ref{fig:fig3} is the
linear least-squares fit to the data (excluding stars
BD~$-13^{\circ}\,3442$ and G~48-29, for which only limits are
available) which is given by $A$(CI)$_{\rm 3D,NLTE}$ -- $A$(CH)$_{\rm
  3D,LTE}$ = 0.483 + 0.240~[Fe/H]$_{\rm 1D,LTE}$ (24 points, with RMS =
0.17) and which we take as the in-principal improvement necessary to
$A$(CH)$_{\rm 3D,LTE}$ to correct it to $A$(CH)$_{\rm 3D,NLTE}$.

That said, given the weakness of the CH features and C~I lines in
metal-poor dwarfs with [Fe/H] $<$ --3.0 (see \citealp{jacobson15},
\citealp{placco16}, and \citealp{fabbian09}), in what follows we shall
assume that this correction is not well-determined below [Fe/H]
{\ltsima} --3.0, and make the conservative assumption that
$A$(CI)$_{\rm 3D,NLTE}$ -- $A$(CH)$_{\rm 3D,LTE}$ = 0.0, for all
[Fe/H], and hence $A$(CH)$_{\rm 3D,NLTE}$ = $A$(CH)$_{\rm 3D,LTE}$.
The reader should bear in mind that the 3D,NLTE corrections we shall
present in what follows are most likely less extreme than would be
obtained by adoption of the equation in the previous paragraph.

\section{Revised $A$(C) vs. [Fe/H] and [C/Fe] vs. [Fe/H] Diagrams} \label{sec:revisions}

\subsection{\it The \citet{yoon16} sample of CEMP stars} \label{sec:yoon}

With the 3D--1D and NLTE--LTE corrections in hand, we now investigate
their effect on the distribution of the CEMP-no stars in the ($A$(C),
[Fe/H]) and ([C/Fe], [Fe/H]) $-$ planes.  Figure~\ref{fig:fig4}
presents the data for the CEMP-no and CEMP-s objects compiled by
\citet{yoon16}\footnote{We used only stars in the \citet{yoon16}
  catalog for which the [C/Fe] values were based on the CH G-band.
  While this does not effect the CEMP-no stars, it excludes some 15
  CEMP-s objects.  We also required the presence of [Ba/Fe] to
  identify membership of the CEMP-no and CEMP-s subclasses, except for
  stars with [Fe/H] $<$ $-$3.3, where in the absence of detected
  [Ba/Fe] we assume CEMP-no status.}, together with those for an
additional recently reported eight CEMP-no stars presented in
Table~\ref{tab:tab4}, and four stars identified in
Table~\ref{tab:tab6}.  The effects of the corrections to $A$(C),
[C/Fe], and [Fe/H] are presented in the three rows of the figure.  The
uppermost row, panels (a) and (b), show data obtained using {\odlte}
for $A$(C)$_{\rm 1D,LTE}$ and [C/Fe]$_{\rm 1D,LTE}$ vs. [Fe/H]$_{\rm
  1D,LTE}$, respectively.  Also shown in the top left panel (a) are
the the ellipses containing the Groups I, II, and III of
\citet{yoon16}, together with the ``high-carbon band'' (horizontal
orange line) and the ``low-carbon band'' (horizontal light blue line)
of \citet{spite13}, \citet{bonifacio15,bonifacio18}, and
\citet{caffau18} (truncated on the right by the [C/Fe] = +1.0
locus)\footnote{We draw the reader's attention to the fact that the
  \citet{yoon16} model contains three components, while that of
  \citet{caffau18} has only two.  The question one might ask is how
  many components are required to best describe the stellar
  distribution.  We shall discuss this further in
  Sections~\ref{sec:namgal} and \ref{sec:toymod}.}.  In these panels,
(a) and (b), red and grey symbols refer to CEMP-no stars on the one
hand, and CEMP-s stars on the other, and in what follows in the middle
and bottom rows the symbol for each star will retain the same color as
adopted in these uppermost panels.  The full and dotted lines in all
panels represent the loci of the [C/Fe] $>$ +0.7 divide between
C-normal and CEMP stars, and [C/Fe] = 0.0, respectively.

In the middle row, panels (c) and (d), we show the effect of
(3D--1D),LTE corrections.  In the left panel (c), the $A$(C)
distribution lies well below that of the 1D,LTE data, and at lower
[Fe/H], than in panel (a).  Here the abscissa becomes [Fe/H]$_{\rm
  3D,LTE}$ in both panels, while the ordinate is $A$(C)$_{\rm 3D,LTE}$
on the left and becomes [C/Fe]$_{\rm 3D,LTE}$ = [C/H]$_{\rm 3D,LTE}$
$-$ [Fe/H]$_{\rm 3D,LTE}$ on the right.  In both panels one sees that
the distribution moves to lower values of [Fe/H], while the carbon
abundances also decrease.  With these corrections, a significant
number of stars now fall below the CEMP limit of [C/Fe] $>$ +0.7: the
number of CEMP-no stars has reduced from 130 in the top row to 68 in
the middle row, a decrease of 48\%.

The bottom row of Figure~\ref{fig:fig4}, panels (e) and (f), presents
the changes when 3D,NLTE corrections are applied to the 1D,LTE data in
the top row. To produce [Fe/H]$_{\rm 3D,NLTE}$ we use the 3D,NLTE
corrections of Section~\ref{sec:fe_subsec2}, and for $A$(C)$_{\rm
  3D,NLTE}$ and [C/Fe]$_{\rm 3D,NLTE}$ adopt corrections following the
discussion in Section~\ref{sec:ci}.  In the context of carbon 3D,NLTE
effects, we recall that in Section~\ref{sec:ci} a comparison of carbon
abundances derived from the CH G~band and from high-excitation C~I
lines leads to the conclusion that the (currently unknowable) NLTE
effects on the G~band appear not to increase CH based abundances (and
indeed hint that they will further decrease them; see
Figure~\ref{fig:fig3}).  Given the the extreme weakness of the CH and
C~I features and the sparseness of data for stars with [Fe/H] $<$
--3.0, we choose to assume $A$(C)$_{\rm 3D,NLTE}$ = $A$(C)$_{\rm 3D,LTE}$
for the purposes of the present discussion.

Inspection of the bottom panels (e) and (f) of Figure~\ref{fig:fig4}
shows that 3D,NLTE considerations have an enormous effect on the
number of putative C-rich stars (exclusively on those in the Yoon
Group II, which have lower $A$(C)$_{\rm 1D,LTE}$).  The number of
CEMP-no stars having [C/Fe] $>$ +0.7 and [Fe/H] $<$ --2.0 in the upper
left panel (a) has decreased from 130 to 35 in the bottom left panel
(e) $-$ a reduction of 73\%.  Against this background it is worth
noting that it is the number of CEMP-no, Group II stars that has
decreased; the number of CEMP-no, Group III stars, which are
considerably more carbon rich, is not affected.

Said differently, 3D,NLTE effects constitute the perfect storm for
those who might wish to understand the carbon abundances of metal-poor
stars by adopting results based on the 1D,LTE assumptions.  First,
1D,LTE overestimates CH-based carbon abundances and, second, it {\it
  underestimates} iron abundances (determined from Fe~I lines) relative
to those determined using 3D,NLTE.  Both effects inflate [C/Fe], and
in both the [C/Fe] vs. [Fe/H], and the $A$(C) vs. [Fe/H] planes the
effects of the transformations are huge. A third effect is that in
these planes the C-rich stars are seen against a considerably larger
C-normal population, in which errors of measurement in both estimates
of, say, 0.2 $-$0.3~dex have the potential to move C-normal stars into
the sparse [C/Fe]-rich region (i.e., into that of the Group II stars).

It has been suggested to the authors that the above results may be
affected by the strong {\logg} sensitivity of the Placco [C/Fe]
corrections present in the \citet{yoon16} data compilation.  That is,
while the corrections are negligible for main sequence stars, they are
large ($\sim$+0.5 dex for objects towards the top of the giant
branch).  Examination of the Placco corrections for [Fe/H] = --3.0 (a
representative value for the present discussion) against the [Fe/H] =
--3.0, age = 12 Gyr isochrone of \citet{demarque04} shows that only
above {\logg} = 2.0 do they become larger than $\sim$0.05.  When we
then replot our Figure~\ref{fig:fig4} including only stars having log
g $<$ 2.0, we find that the areas covered by the stars are not
significantly changed from the point of view of the present discussion
of the 3D and NLTE predictions.  In particular, the number of CEMP-no
stars on the 1D,LTE panel is 66, which reduces to 40 for 3D,LTE, and
24 for 3D,NLTE -- reductions of 39 and 64\%, respectively (compared
with 48\% and 73\% for the complete sample).
 
\subsection{\it The \citet{yong13a} sample of CEMP and C-normal stars} \label{sec:yong13a}

We have also applied the above formalism to the literature sample of
190 extremely metal-poor stars of \citet{yong13a}.  An advantage of
this sample is that it is not limited to only CEMP stars as is that of
\citet{yoon16}.  It was used by \citet{yong13b} to place constraints
on the MDF of metal-poor stars, and on the fraction of C-rich stars as
a function of metallicity ([Fe/H]). Re-examination of the
\citet{yong13a} sample has the potential to highlight the effects that
the correction of abundances from 1D,LTE to 3D,LTE and 3D,NLTE has on
these important relationships, not only on the C-rich stars but also
on those that are C-normal.

Figure~\ref{fig:fig5} presents $A$(C) and [C/Fe] as a function of
[Fe/H]\footnote{The carbon abundances have been corrected for
  evolutionary effects following \citet{placco14}.}, where the layout
in the figure is the same as that of Figure~\ref{fig:fig4} and we have
adopted the same 3D,LTE and 3D,NLTE corrections. The red symbols refer
to CEMP-no stars that have carbon abundances based on detections
yielding [C/Fe]$_{\rm 1D,LTE}$ $>$ +0.7 and [Ba/Fe]$_{\rm 1D,LTE}$ $<$
0.0; the green symbols represent C-normal stars with either carbon
detections or limits having [C/Fe]$_{\rm 1D,LTE}$ $<$ +0.7; and grey
circles stand for CEMP-s stars\footnote{As in Section~\ref{sec:yoon}
  we have excluded from our analysis stars having carbon abundances
  based on the C$_{2}$ molecule}.  As in our discussion of
Figure~\ref{fig:fig4}, in the middle and bottom rows the symbol for
each star retains the same color as adopted in the upper, 1D,LTE,
panels.  For 1D,LTE-based abundances and [Fe/H]$_{\rm 1D,LTE}$ $<$
$-$2.0, there are some 88 C-normal and 28 CEMP-no stars.  CEMP-no
stars represent a fraction of 24\% of the total of CEMP-no plus
C-normal stars.

Figure~\ref{fig:fig5} also permits an estimate of the 3D and NLTE
effects on the fraction of CEMP-no stars in a sample containing both
C-normal and CEMP-no stars. As in Figure~\ref{fig:fig4}, the middle
and bottom rows refer to abundances determined assuming 3D,LTE and
3D,NLTE, respectively, and here too large fractions of 1D,LTE CEMP-no
stars become C-normal when investigated in 3D and NLTE.  For stars
with [Fe/H]$_{\rm 3D,LTE}$ $<$ $-$2.0 in the bottom row, $A$(C)$_{\rm
  3D,NLTE}$ vs. [Fe/H]$_{\rm 3D,NLTE}$, there are some 108 C-normal
and 9 CEMP-no stars, leading to a fraction of CEMP-no stars of 8\%, a
very significant decrease compared with the 1D,LTE fraction
of 24\%. (As noted in the previous section, it is the
number of Group II stars that is decreasing, while that of their Group
III counterparts remains unchanged.)

A somewhat surprising result evident in the bottom panels is the
number of stars well below [C/Fe]$_{\rm 3D,NLTE}$ {\simlt} 0.0.  For
these stars, and ignoring those having only carbon abundance limits,
$\langle$[C/Fe]$_{\rm 3D,NLTE}$$\rangle$ = --0.42 $\pm$ 0.03, with
dispersion $\sigma$ = 0.27 (82 objects). In comparison, for dwarfs
with --3.2 $<$ [Fe/H] $<$ --2.0, \citet[see their Figure 1]{amarsi19},
from analysis of the infrared high-excitation C~I lines, report
[C/Fe]$_{\rm 3D,NLTE}$ values $\sim$~+0.1 dex. While a full
explanation of the present G-band abundances lies outside the scope of
the present work, we make two comments.  The first is that the effect
appears to be gravity dependent, insofar as for giants ({\logg} $<$
3.35) in the present sample we find $\langle$[C/Fe]$_{\rm
  3D,NLTE}$$\rangle$ = --0.49 $\pm$ 0.04 (61 stars) and for dwarfs
({\logg} $>$ 3.35) $\langle$[C/Fe]$_{\rm 3D,NLTE}$$\rangle$ = --0.24
$\pm$ 0.03 (21 stars).  Further support for the higher value obtained
for dwarfs is provided by the data for those in our Table 3.  For the
stars in the table with CH-based carbon abundances determined in the
present work, and excluding stars with only limits, we find
$\langle$[C/Fe]$_{\rm 3D,NLTE}$$\rangle$ = --0.18 $\pm$ 0.03. A
possible explanation of the effect is that the giant abundances have
been underestimated.  The second point is that \citet{gallagher16}
have reported that the $A$(C) 3D,LTE corrections are a function of
$A$(C) (see our Section~\ref{sec:uncertainties}).

\subsection{\it A comment on the CEMP-no status of {\cdsean}, G64-12, and G64-37}\label{sec:ch3d1d} 

{\cdsean}, G~64-12, and G~64-37 were recently re-classified as CEMP-no
stars by \citet{jacobson15} and \citet{placco16}. They are all
extremely metal-poor near main-sequence-turnoff stars with similar
{\teff}, {\logg}, [Fe/H]$_{\rm 1D,LTE}$ (--3.41, --3.29, --3.11), and
[C/Fe]$_{\rm 1D,LTE}$ (1.10, 1.07, and 1.12), together with [Ba/Fe] =
$<$--1.05.  --0.36, and --0.06, respectively.  The 3D,LTE and 3D,NLTE
iron and carbon abundances of these objects, however, argue that all
of them are C-normal, with an average carbon abundance of
$\langle$[C/Fe]$_{\rm 3D,LTE}$$\rangle$ = 0.74 and
$\langle$[C/Fe]$_{\rm 3D,NLTE}$$\rangle$ = 0.26.  Some support for
this conclusion is suggested by the fact that their discovery as
metal-poor stars is based on their halo kinematics (\citealp{carney81}
and \citealp{ryan91}), without knowledge concerning their abundance
characteristics, i.e., they are an unbiased sample with respect to
abundance\footnote{We implicitly assume that halo stars chosen by
  their extreme kinematics are drawn without bias from the same
  population as halo stars selected by their extreme metal
  deficiency.}.  If one accepts that the CEMP fraction at [Fe/H] =
--3.2 is $\sim$~0.30, based on 1D,LTE analyses (\citealp{yong13b},
\citealp{lee13}), the probability that all three of them should be
CEMP stars is only $\sim$~3\%.

\subsection{\it On the Nature of the Group I CEMP-no Stars} \label{sec:group1_cempno}

One of the most intriguing aspects of the \citet{yoon16} $A$(C) vs.
[Fe/H] diagram is that while all of the CEMP-s stars belong to Group I
and their Groups II and III contain only CEMP-no stars, there is a
non-negligible fraction of CEMP-no stars within the Group I boundary.
Inspection of our Figure~\ref{fig:fig4} shows there are 19 such
CEMP-no Group I stars\footnote{Guided by Figure~\ref{fig:fig4}, we
  required the CEMP-no stars to have $A$(C) $>$ 7.1 and --3.9 $<$
  [Fe/H] $<$ --2.0.},which represent 14\% of the group.  The obvious
question is: why is the abundance of carbon relative to hydrogen
higher by some $\sim$1 dex in the CEMP-no, Group I stars compared with
that of the majority of their counterparts in Groups II and III?  For
future reference, Table~\ref{tab:tab5} presents details of the 19
CEMP-no, Group I stars.

To our knowledge, the significance of this subset of CEMP-no stars was
first appreciated by \citet{hansent16}, who reported that five of
their sample of 24 CEMP-no stars (HE~0219--1739, HE~1133--0555,
HE~1410+0213, HE~1150--0428, and CS~22957--027) lie in or close to the
``high-carbon band'' first reported by \citet{spite13} (in large part
their ``high-carbon band'' is related to the \citet{yoon16} Group I).
\citet{hansent16} noted that three of these are binaries.  They
commented: ``Should the majority [of this subset of CEMP-no stars]
turn out to be members of binary systems ... and in particular if
there are signs that mass transfer has occurred, this would lend
support to the existence of AGB stars that produce little if any
s-process elements''.  \citet{arentsen18} have further addressed the
issue and reported that this CEMP-no subset has a ``binary fraction
... of $47^{\,+15\,}_{\,-14}\%$ for stars with higher absolute carbon
abundance''.  Inspection of our Table~\ref{fig:fig5} shows that four
out of nine CEMP-no, Group I (i.e., $\sim$~45\%) stars for which data
are available are binary.

A further related conjecture is that the putative AGB stars of Hansen
et al. may have been the 7~M$_\odot$, initial rotational velocity 800
km s$^{-1}$ spinstars of \citet{meynet06}, which very nicely
reproduce the light-element abundance patterns of the CEMP-no stars.
Taking the potential binary-with-mass-transfer hypothesis further, if
one assumes that the number of Group I CEMP-s stars is proportional to
the number of putative polluting stars in the mass range (say) 2 --
6~M$_\odot$ (cf., \citealp{lugaro12}), while the number of Group I
CEMP-no is proportional to that of those in the mass range (say) 6 --
8~M$_\odot$, (e.g., \citealp{meynet06}), and further assumes that star
formation followed the Salpeter Initial Mass Function, one finds that
the ratio of AGB stars in the 6 -- 8~M$_\odot$ range to those in the
two mass ranges together is 0.09\footnote{The fraction is somewhat
  sensitive to the lower mass limit of the low mass range.  Had we
  chosen 1 -- 6~M$_\odot$, or 3 -- 6~M$_\odot$, the fraction would
  have changed from 0.09 to 0.03, or 0.17, respectively.}, similar to
the observed fraction, 0.14, noted above, which CEMP-no, Group I stars
contribute to the total Group I sample.

\section {The Metallicity Distribution Function (MDF) and the CEMP-no Fraction} \label{sec:yong13b}

The MDF of the Galaxy's very metal-poor stars (VMP, [Fe/H] $<$ $-$2.0)
is complicated by the fact that the sample is inhomogeneous,
comprising several sub-populations.  A topic closely related to the
MDF is the size of the CEMP-no fraction, and its dependence on metal
abundance.  Any understanding of these will ultimately turn on a
closer knowledge of the metallicity distribution functions of the
halo's several components.  In this context, the two major C-rich
groups, of CEMP-s and CEMP-no stars, provide an interesting challenge.
We refer the reader to Papers III and IV of this series
(\citealp{yong13b}, \citealp{norris13}), and references therein, for
an effort to better understand the role of these sub-populations.
Other important investigations include those of \citet{carollo12,
  carollo14} and \citet{lee13,lee17}.  The question we shall address
here is the role that 3D,NLTE corrections to 1D,LTE carbon and iron
abundances play in our understanding of these matters.

\subsection{\it MDFs} \label{sec:mdfs}

We use the formalism of \citet{yong13b} to examine the MDF of the
\citet{yong13a} sample discussed in the previous section.  C-rich
stars ([C/Fe] $>$ +0.7) were included only if an abundance was
available in \citet{yong13a} (i.e., those with an abundance limit were
excluded), while both detections and limits were included in the
C-normal regime ([C/Fe] $<$ +0.7). In the left panel of
Figure~\ref{fig:fig6}, the logarithm (base 10) of the generalized
histogram (adopting a Gaussian kernel having $\sigma$ = 0.30~dex) is
presented as a function of [Fe/H], where [Fe/H] is adopted as proxy
for the total heavy element abundance.  As in our earlier work, we
investigate the MDF for stars with [Fe/H] $<$ $-$3.0, and to which we
have applied sample completeness corrections.  In the figure, the
green-shaded areas pertain to the combination of the CEMP-no and
CEMP-s subgroups, the grey-shaded areas refer to C-normal stars, and
the small unshaded (upper) areas stand for stars for which the carbon
abundance was not measured.

The top panel of the figure is based on 1D,LTE abundances, while the
middle and bottom panels present 3D,LTE and 3D,NLTE results,
respectively.  The outstanding feature of the MDFs is the decreasing
role of the C-rich stars ([C/Fe] $>$ +0.7) when 3D and non-LTE
corrections are applied, as would be expected from inspection of
Figure~\ref{fig:fig5}.

\subsection{\it CEMP-no fraction} \label{sec:fraction}

In the right panel of Figure~\ref{fig:fig6} we present the manner in
which the fraction of CEMP-no stars increases as [Fe/H] decreases when
one changes from 1D,LTE to 3D,LTE, to 3D,NLTE.  Here we define the
CEMP-no fraction as N$_{\rm CEMP-no}$/(N$_{\rm C-normal}$ + N$_{\rm
  CEMP-no}$ + N$_{\rm CEMP-s}$), where we include the CEMP-s stars in
the equation on the assumption they were once C-normal stars, in order
to obtain a more complete fraction.  In practice, this has only a
small effect in the present discussion, given there are relatively few
CEMP-s stars with [Fe/H] $<$ $-$3.0.

In this panel, the full lines represent the fraction of CEMP-no stars,
while for comparison purposes the dashed line in each of the lower two
subpanels is the 1D,LTE fraction presented in the topmost subpanel.
In the top, middle, and bottom panels the fraction of CEMP-no stars
with --4.5 $<$ Fe/H] $<$ --3.0 are 0.30, 0.15, and 0.12, respectively.
We also note that while in this figure all stars with [Fe/H] $<$ --4.5
are C-rich, the relatively complete sample upon which this is based
contains only three such objects.  We recall from our
Tables~\ref{tab:tab1} and \ref{tab:tab6} that, at time of writing,
some 14 stars are now known with [Fe/H] $<$ --4.5, a large majority of
which is C-rich (see \citealp{fn15}, \citealp{frebel15},
\citealp{caffau16}, \citealp{aguado_2}, \citealp{aguado_1}, and
\citealp{starkenburg18}).  In the present context, perhaps the most
significant result one might take from the panel is that when one
includes 3D and NLTE corrections a separation between the population
of C-rich stars with [Fe/H] $<$ --4.5, and that with [Fe/H] $>$ --4.5
becomes clearer, and more significant.

\section{Uncertainties} \label{sec:uncertainties}

We alert the reader to some uncertainties implicit in the present work.  

\subsection{\it The CH 3D,LTE corrections are a function of $A$(C)}

\citet{gallagher16} report that G~band 3D,LTE corrections are a
function not only of [Fe/H], but also of $A$(C), and present a
comprehensive investigation of 3D corrections for CEMP dwarfs on the
ranges $-$3.0 $<$ [Fe/H] $<$ $-$1.0, 5900K $<$ {\teff} $<$ 6500K, 4.0
$<$ {\logg} $<$ 4.5, 6.0 $<$ $A$(C)$_{\rm 3D}$ $<$ 8.5, and for two
values of C/O = 0.21 and 3.98.  They emphasize that the corrections
are sensitive to the C/O ratio, and adopt the value of C/O = 0.21 as
most appropriate for the CEMP-no stars.  We note here that for CEMP-no
stars the available abundance data suggest that [O/Fe] increases
linearly with [C/Fe] (e.g., \citealp[Figure 2]{norris13}), and
therefore a constant value of C/O.

\subsection{\it How trustworthy is the Drawin scaling factor S$_{\rm H}$ treatment of the neutral hydrogen?}  

In Section~\ref{sec:ci}, the carbon abundances based on the analysis
of high-excitation C~I lines adopted the formalism of Drawin to
describe the influence of inelastic hydrogen atom collisions.
\citet{barklem11}, however, report that ``Quantitatively, the Drawin
formula compares poorly with the results of the available quantum
mechanical calculations, usually significantly overestimating the
collision rates by amounts that vary markedly between transitions.''
That said, we recall here, from Section~\ref{sec:ci}, the excellent
agreement between the analyses of \citet{fabbian09} and
\citet{amarsi19}, the latter of which adopts ``modern descriptions of
the inelastic collisions with neutral hydrogen''.

\subsection{\it Are differences between photometric and spectroscopic {\teff} values a problem?}

An evergreen uncertainly in the determination of chemical abundances
based on 1D,LTE analyses is the differences that result when {\teff}
values are based on different assumptions.  Suffice it here to say
that [Fe~I/H]$_{\rm 1D,LTE}$ values can differ systematically by
values of order 0.3 -- 0.4~dex between analyses that adopt photometric
{\teff} values and those that use spectroscopically determined ones
(see e.g., \citealp[Table 17]{roederer14}).  This could be important
in determining Group II, CEMP-no status, for example, in
Figures~\ref{fig:fig4} and \ref{fig:fig5}.

\section{The Abundances of Other Elements in the CEMP-no Stars}\label{sec:naba}

\subsection{\it The Light Elements Na, Mg, and Al}\label{sec:namgal}

A distinctive feature of the CEMP-no stars is that they also exhihit
overabundances of Na, Mg, and Al, to varying degrees in size and from
element to element, while only small (if any) differences are found in
the relative abundances, [X/Fe], on the range Si through to the
heavy-neutron-capture elements (see \citealp{fn15}, and references
therein).  Indeed, the interpretation of [X/Fe] as a function of
atomic number is a key to an understanding of the origin of these
stars (see \citealp{umeda03}, \citealp{meynet06}, \citealp{heger10},
\citealp{nomoto13}, \citealp{takahashi14}, \citealp{maeder15b}, and
references therein).  That said, insofar as discussed in
Section~\ref{sec:revisions}, many stars which under the 1D,LTE
assumption were designated CEMP-no become C-normal when interpreted
using 3D,NLTE, it is probably fair to say that we may not yet have a
complete understanding of these abundance patterns.  A potential
example of this problem is the report by \citet{yoon16} that their
Group II and Group III CEMP-no stars have different Na and Mg
distributions.  A second interesting phenomenon, described in
Section~\ref{sec:group1_cempno}, is the existence of a $\sim$15\%
subpopulation of CEMP-no stars in their Group I, which principally
comprises only CEMP-s stars.  What is the light element signature of
these stars?

Figure~\ref{fig:fig7} shows generalized histograms of 1D,LTE
abundances for [Na/Fe], [Mg/Fe], and [Al/Fe], together with
[Ca/Fe]\footnote{Figures~\ref{fig:fig7} and ~\ref{fig:fig8} are based
  on data from \citet{aoki13}, \citet{bonifacio15}, \citet{barklem05},
  \citet{christlieb04}, \citet{cohen08, cohen13},
  \citet{frebel14,frebel15}, \citet{hansent15,hansenc16},
  \citet{hollek11}, \citet{ito13}, \citet{jacobson15},
  \citet{norris10,norris13}, \citet{placco14,placco16},
  \citet{plez05}, \citet{roederer14}, \citet{spite18}, \citet{yong13a},
  and \citet{yoon16}.} (which exhibits small if any variation, and is
shown here only for comparison purposes). In each sub-panel the
thicker red line pertains to CEMP-no stars, and the thinner black one
to C-normal objects; and the areas under both curves have been
normalized to be the same. In the top three rows of the figure,
results are presented for the Group I and III CEMP-no stars, where
from top towards bottom we consider Group III, Group I, and Group III
+ Group I CEMP-no stars, respectively.  (For the C-normal stars the
same histogram is presented in all sub-panels of the same element.)
The numbers of CEMP-no stars involved are presented in each sub-panel,
and while they are small, one's first impression is that the Na, Mg,
and Al distributions are similar in both Groups I and III, at least
insofar as significant overabundances are evident in all panels; and
while more data are required, the figure suggests that Group I and
Group III CEMP-no stars have experienced similar enrichment pathways.

The bottom two rows present abundances for CEMP-no, Group II stars.
The upper of the two rows is for [C/Fe]$_{\rm 1D,LTE}$ $>$ +1.0, while
the lower is the result for +0.7 $<$ [C/Fe]$_{\rm 1D,LTE}$ $<$ +1.0.
Inspection of the two panels suggests a broader distribution of each
of Na, Mg, and Al abundances (but not of Ca) in the upper row than in
the lower one.  The simplest interpretation of this difference is that
the majority of the stars in the range +0.7 $<$ [C/Fe]$_{\rm 1D,LTE}$
$<$ +1.0 does not have overabundances of Na, Mg, and Al, or that their
carbon abundances have been overestimated.  This could also explain,
at lease in part, the report by \citet{yoon16} that the Group II and
III CEMP-no stars have experienced different Na and Mg enrichment
pathways.  
 
It has been suggested to us that the \citet{yoon16} Group II stars are
not CEMP-no stars, and have apparently large carbon abundances due to
errors of measurement, and/or of the \citet{placco14} corrections.
This is not obvious to us, given that the Group II stars with [C/Fe]
$>$ +1.0 are identified as CEMP-no by both \citet{yoon16} and
\citet{caffau18} (see Figure~\ref{fig:fig4}), and that some of them
have Na, Mg, and Al overabundances.  Further work is needed to address
this issue.  The outstanding question for us is: why is the
distribution of CEMP-no stars in Figure~\ref{fig:fig4} so obviously
non-uniform, leading \citet{yoon16} to identify two groups.  We shall
return to this in Section~\ref{sec:toymod}.

We conclude our discussion by noting the enigmatic result that while
overabundances of Na and Mg in the CEMP-no stars are clear in all
panels which present these elements in Figure~\ref{fig:fig7} (except
for Mg in the bottom row), the histograms also appear to have a
component that has close to the solar abundance ratio. More data are
clearly required to confirm and address the reality and implications
of this effect.

\subsection{\it [Sr/Ba] and the nature of the CEMP-no, Group I stars}\label{sec:srba}

How may one understand the CEMP-no, Group I stars.  In
Section~\ref{sec:group1_cempno} we noted the suggestion of
\citet{hansent16}, supported by further work of \citet{arentsen18},
that the binarity of a significant fraction of these stars might
signal mass transfer in a system in which the AGB star did not
experience s-process enhancement.  We also pointed out that the ratio
of Group I CEMP-no to CEMP-s stars is consistent with higher masses
for the putative AGB star enrichment of the CEMP-no, Group I stars
than exists for their CEMP-s, Group I counterparts.

The question then is, do descriptions of massive AGB stars that
produce primary carbon, but little or no s-process enhanced material,
exist in the literature?  The obvious answer is the extremely
metal-poor spinstars of \citet{meynet06} and \citet{frisch10,
  frisch12}, which do not produce the s-process pattern, but rather
overproduce Sr relative to Ba.  In this context, Hansen et al. (2019)
have proposed [Sr/Ba] as a parameter to distinguish between the
various CEMP subclasses, based in part on the result of Frischknecht
et al. that Sr is overproduced more relative to Ba than is observed in the
CEMP-no and C-normal stars. 

Against this background, Figure~\ref{fig:fig8} presents [Sr/Ba]$_{\rm
  1D,LTE}$ as a function of [Fe/H]$_{\rm 1D,LTE}$ for CEMP-no stars of
Group I (red star symbols) and III (red circles), together with CEMP-s
stars (grey circles), based on data from the literature.  The
important result here is that, taken as a whole, the [Sr/Ba] values of
the CEMP-no, Group I objects are larger than those of the CEMP-s
stars{\footnote{An exception to this rule, not included in the
    \citet{yoon16} compilation, is SDSS J0222--0313, which has [Fe/H]
    = --2.65 and [Sr/Ba] = 1.02 \citep{caffau18}.}}.  We also note
that most of the [Sr/Ba] values for the Group III, CEMP-no stars are
lower limits, and that their values could be as large as those of the
Group I, CEMP-no stars.  One might envisage scenarios involving
spinstars and/or binarity.

\section {A Toy Model for the CEMP-no Stars in the $A$(C), [C/Fe] vs. [Fe/H] Planes}\label{sec:toymod}

A fundamental problem in understanding the formation of the first
stellar populations is the manner in which the initial gas clouds
cooled to form stars.  We refer the reader to \citet{frebel07},
\citet{schneider12}, \citet{chiaki17}, and references therein, for
details on the role of the various cooling mechanisms and pathways in
which this may have proceeded.  We present here a very simple toy
model that seeks to explain the distribution of CEMP-no stars in the
$A$(C), [C/Fe] vs. [Fe/H] planes in the first few hundred Myr.

We proceed with the following set of assumptions:{\vspace{-4mm}

\begin{itemize}

\item
The first generation of stars produced an initially carbon rich
environment in which further star formation proceeded along two
principal pathways, one forming extremely carbon rich objects (seen
today as the C-rich stars with [Fe/H] {\simlt } --4.5 -- the minority
population), the other (later) one comprising C-normal stars (seen
today as the bulk of stars with [Fe/H] {\simgt} --4.0 -- the majority
population).

\item
CEMP-no stars with [Fe/H] {\simgt} --4.0 formed following the
coalescence of gas clouds of these C-rich and C-normal populations.

\item
Our basic toy model assumption is that in each coalescence of C-rich
and C-normal gas clouds their individual masses are determined by the
mass function of the respective populations, which we shall assume to
be the Salpeter mass function.  This mass function is, of course,
determined from the observation of stars, rather than of gas clouds;
but that said, support for adopting a power-law mass function for the
clouds has been reported by \citet{elmegreen02}.  We further assume
that the mass of the putative composite star is the sum of those of
the two gas clouds.  In order to proceed, we draw masses at random
from the Salpeter mass function on the range 0.10 $\leq$ M/M$_{\odot}$
$\leq$ 0.75 for each of the carbon classes and accept a composite star
if the sum of masses lies in the range 0.65 $\leq$ M/M$_{\odot}$
$\leq$ 0.75, which approximately covers that observed for the
metal-poor stars with [Fe/H] $<$ --2.0 discussed here.

\item
We determine chemical abundances ([Fe/H], $A$(C), and [C/Fe]), on the range
[Fe/H] $<$ --2.0, as follows.  For C-normal stars we adopt the MDF of
\citet{yong13b} (transformed to the [Fe/H]$_{\rm 3D,NLTE}$ scale), and
draw [Fe/H] at random from that distribution, and assume [C/Fe] = 0.0
(close to the 3D,NLTE values obtained by \citealp{fabbian09} and
\citealp{amarsi19}).  For the C-rich population, here defined to lie
in the range [Fe/H] $<$ --4.5, and for which we have little
information on the MDF, we assume individual values of [Fe/H] and
[C/Fe] suggested by the observed values of the 10 C-rich stars for
which we have information (i.e., in the ranges --6.0 {\simlt} [Fe/H]
{\simlt} --4.5, and +1.0 {\simlt} [C/Fe] {\simlt} +5.0).

\end{itemize}

Using these concepts we attempt to learn only about the regions of the
$A$(C) and [C/Fe] vs. [Fe/H] planes that are occupied by the putative
composite stars, and emphasize that these results should be seen as a
zeroth-order approximation.  The first assumption is that merging
occurs between between clouds of random mass (say M$_{\rm C-rich}$ and
M$_{\rm C-normal}$) adopting a Salpter mass function, one from each of
the two parent populations, to form a composite CEMP-no star in the
currently observable mass range 0.65 -- 0.75 M/M$_{\odot}$.  We
implicitly assume that there are reservoirs of gas having the required
masses in the two parent populations, and that all of the merging
material is used to produce a well-mixed composite star, without mass
loss.  We then determine the chemical abundances of carbon and iron of
each of the coalescing gas clouds.  For the C-rich component we adopt
a representative pair of values (e.g., [Fe/H]$_{\rm 3D,NLTE}$ = --5.0,
and [C/Fe]$_{\rm 3D,NLTE}$ = +3.0).  For the cloud from the C-normal
population we determine [Fe/H]$_{\rm 3D,NLTE}$ at random from the
modified \citet{yong13b} MDF over the range --4.0 $<$ [Fe/H]$_{\rm
  3D,NLTE}$ $<$ --2.0, and assume [C/Fe]$_{\rm 3D,NLTE}$ = 0.0.  Given
M$_{\rm C-rich}$ and M$_{\rm C-normal}$, and these chemical abundances
for the two clouds, the abundances of the composite star follow.  We
emphasize that with this approach we seek only to reproduce the
observations of the CEMP-no stars with [Fe/H] $>$ --4.0 in
Figure~\ref{fig:fig4}.  The results of a number of simulations are
shown in Figures~\ref{fig:fig9} -- \ref{fig:fig11}.

Figure~\ref{fig:fig9} pertains to a C-rich parent population having
[Fe/H] = --5.0 and [C/Fe] = +3.5, which coalesces with C-normal halo
clouds as postulated above. On the left are the computed model
results, labeled 3D,NLTE on the assumption that 3D,NLTE observational
data would re-produce these results.  On the right, labeled 1D,LTE,
are the results when the model values are reverse engineered to
produce values that would be obtained by a 1D,LTE analysis.  The star
symbols refer to the adopted values of the C-rich population, while
the small symbols represent the composite model results.
Figure~\ref{fig:fig10} presents a considerably smaller carbon
abundance of the C-rich parent population, with [Fe/H] = --5.0 and
[C/Fe] = +1.5, which lead to considerably different abundance
distributions compared with the simulation in Figure~\ref{fig:fig9}.

In Section~\ref{sec:namgal}, we noted that the distribution of CEMP-no
stars in Figure 4 is obviously non-uniform, leading Yoon et al.
(2016) to identify two groups of CEMP-no stars.  In
Figure~\ref{fig:fig11}, we present a comparison of our model results,
which contains three simulations and their coaddition, with the
\citet{yoon16} Groups I, II, and III boundaries -- which we recall
were defined in this plane.  In the figure, simulated $A$(C)$_{\rm
  1D,LTE}$ (left) and [C/Fe]$_{\rm 1D,LTE}$ (right) values are plotted
as a function of [Fe/H]$_{\rm 1D,LTE}$, where the format is similar to
that of our Figure~\ref{fig:fig4}.  In the upper left of each of the
uppermost three rows in the figure are the assumed [Fe/H]/[C/Fe]
parameters of the C-rich population, and assuming these are also the
3D,NLTE values of that population, the large red star symbols
represent the corresponding 1D,LTE values. The results of the
coalescence model are presented as small red circles, the number of
which is given by the isolated number in each of these panels.

Only one parameter, [C/Fe] of the C-rich population, changes among the
top three rows of the figure -- from +4.5 to +3.0, to +1.5; and at
least to first approximation, one can see a reasonable reproduction of
the Yoon et al. Groups I, III, and II, respectively, in our
Figure~\ref{fig:fig4}, proceeding from top towards bottom. Finally,
the bottom row of the figure presents the simple co-addition of the
data in the upper three rows, and should be compared with the
uppermost row of Figure~\ref{fig:fig4}. 

Perhaps the most interesting feature of the figure is that in the left
column ($A$(C) vs. [Fe/H]) the morphology of the distribution of
coalesced stars in the top panel ([C/Fe]$_{\rm C-rich}$ = +4.5), which
is similar to that of \citet{yoon16} Group III, has changed in the
second panel from the bottom ([C/Fe]$_{\rm C-rich}$ = +1.5) to that of
\citet{yoon16} Group II.  This suggests that the morphology of the
CEMP-no stars in this plane is determined by the distribution of
carbon in the C-rich cloud population.

We regard the agreement between the model and the observational data
as encouraging, given the {\it ad hoc} nature and simplicity of the
assumptions of the model.

\section{Summary and Desiderata} \label{sec:summary}

In Sections~\ref{sec:corrections} -- \ref{sec:uncertainties} we
applied literature-based 3D,NLTE corrections to 1D,LTE Fe~I iron and
CH-based carbon abundances for stars with [Fe/H]$_{\rm 1D,LTE}$ $<$
--2.0, with a view to obtaining a better understanding of the nature
and origin of the CEMP-no stars and what they have to tell us about
the most iron-poor ([Fe/H] $<$ --4.5) C-rich stars and their
relationship to and interaction with the majority iron-poor ([Fe/H]
$>$ --4.0), carbon-normal halo population.  Bootstrapping from carbon
abundances based on 3D,NLTE analysis of the infrared high-excitation
C~I lines in the range --3.3 $<$ [Fe/H] $<$ --2.0, we showed that
although it is not currently possible to theoretically determine NLTE
corrections for the CH molecule, the 3D,NLTE corrections are very
likely not smaller (absolutely) than the 3D,LTE values.  As emphasized
by \citet{asplund05}, the resulting corrections are very large.  For
example, for the \citet{yoon16} compilation of C-rich stars, if one
adopts [C/Fe] $>$ +0.7 as a basic requirement of a CEMP-no star, the
fraction of CEMP-no stars in the range --4.5 $<$ Fe/H] $<$ --3.0 drops
from the 1D,LTE result of 0.30 to the 3D,NLTE value of 0.12.

\begin{itemize}
\item
In Section~\ref{sec:yong13a}, we found a large number of C-normal
stars below [C/Fe]$_{\rm 3D,N LTE}$ {\simlt} 0.0 for which the
(CH-based) mean [C/Fe]$_{\rm 3D,NLTE}$ abundance is
$\langle$[C/Fe]$_{\rm 3D,NLTE}$$\rangle$ = --0.42 (82 objects),
surprisingly low compared with the result of [C/Fe]$_{\rm 3D,NLTE}$
$\sim$~+0.1 based on high-excitation C~I lines in metal-poor dwarfs
reported by \citet{amarsi19}. This might be attributed to the fact
that the present sample is dominated by giants.  That is, for giants
in the present sample we find $\langle$[C/Fe]$_{\rm 3D,NLTE}$$\rangle$
= --0.49 (61 stars) and for dwarfs $\langle$[C/Fe]$_{\rm
  3D,NLTE}$$\rangle$ = --0.24 (21 stars).  Also, analysis of C-normal
dwarfs in our Table~\ref{tab:tab3} finds $\langle$[C/Fe]$_{\rm
  3D,NLTE}$$\rangle$ = --0.18 (21 stars).

Perhaps the 1D,LTE values for the giants have been underestimated.
Alternatively, we noted that \citet{gallagher16} have reported that 3D
CH-based carbon abundances are a function of $A$(C).  While it could be
a massive undertaking, it would be interesting to further investigate
parameter space to better understand this effect.
\end{itemize}

\begin{itemize}
\item
It may be suggested that 1D,LTE [C/Fe] abundances are no longer of
use.  We would counter that this would be premature, and are of the
view that while comprehensive 3D,NLTE investigations of parameter
space are required, further 1D,LTE surveys are important to discover
further objects in order to constrain and calibrate the 3D,NLTE
predictions.
\end{itemize}

It was noted in Section~\ref{sec:revisions} that the change of CEMP-no
status applies to the Yoon et al. Group II, CEMP-no stars, and not to
their C-richer Group III counterparts.  An important example of the
Group II effect of the corrections on 1D, LTE CH based carbon
abundances is the status change of the three Group II classic
metal-poor ([Fe/H] $\sim$~--3.0) stars {\cdsean}, G64-12, and G64-37,
for which $\langle$[C/Fe]$_{\rm 1D,LTE}$$\rangle$ = 1.1 and
$\langle$[C/Fe]$_{\rm 3D,NLTE}$$\rangle$ = 0.3
(Section~\ref{sec:ch3d1d}).  In Section~\ref{sec:namgal}, a second
enigmatic result for Group II objects arose in the discussion of
1D,LTE abundances of Na, Mg, and Al (which show large overabundances
in the CEMP-no stars) where we found for the Group II stars that while
variations in the abundance histograms of these elements were seen in
the abundance range [C/Fe]$_{\rm 1D,LTE}$ $>$ +1.0, the effect is not
so evident for stars with +0.7 $<$ [C/Fe]$_{\rm 1D,LTE}$ $<$ +1.0.

We also discussed the existence of a $\sim$15\% component of CEMP-no
stars in the Yoon et al. Group I of CEMP stars, which is comprised
principally of CEMP-s stars.  In this subgroup of CEMP-no stars we
found that the [Sr/Ba]$_{\rm 1D,LTE}$ values are larger than those of
the majority of the CEMP-s stars.

\begin{itemize}
\item
Further work is needed to investigate to what extent the stars in this
CEMP-no subgroup may be binary and/or the progeny of spinstars.
\end{itemize}

Finally, we presented a toy model that seeks to describe the formation
of CEMP-no stars in the abundance range --4.0 {\simlt}
  [Fe/H] {\simlt}~--2.0 in terms of the coalescence of pre-stellar
clouds of the two populations that followed the chemical enrichment by
the first zero-heavy-element stars, that is, the C-rich,
hyper-metal-poor population and the C-normal, extremely-metal-poor,
halo stars having [Fe/H] {\simgt}~--4.0.  The simplicity of the model,
and the uncertainty of the Fe and C abundance distributions and mass
function of the hyper-metal-poor population notwithstanding, the model
produces abundance behavior in the $A$(C)$_{\rm 1D,LTE}$ and
[C/Fe]$_{\rm 1D,LTE}$ vs. [Fe/H]$_{\rm 1D,LTE}$ planes not unlike that
seen in the \citet{yoon16} Groups I, II, and III.

\begin{itemize}
\item
A more rigorous approach to this simple coalescence model would seem worthwhile.
\end{itemize}

\section{APPENDIX: THE 14 MOST IRON-POOR STARS}

In Table~\ref{tab:tab6}, we present details for the 14 iron-poor stars
currently known to have [Fe/H] $<$ --4.5.  Columns (1) -- (3) contain
starname and coordinates, columns (4) -- (6) present atmospheric
parameters {\teff}, {\logg}, and [Fe/H], column (7) contains [C/Fe],
and source information is presented in the final column.  In this
table the abundances assume 1D,LTE.

\acknowledgements

We are grateful to the referee, Piercarlo Bonifacio, for a comprehensive report, which led
to significant improvements to the manuscript.  We thank Martin
Asplund for enlightening discussions on 3D,NLTE matters over the past
two decades, and Thomas Nordlander for his very helpful advice during
the present investigation.  We thank Timothy Beers for his insightful
discussions on carbon rich stars over many years, and his helpful
reading of the manuscript.  Vini Placco kindly provided [C/Fe]
evolutionary corrections adopted in Section~\ref{sec:yong13a}.
J.E.N. acknowledges advice from his Ph.D.  supervisor Leonard Searle:
``You may be diffident about being that bold, but it's better to be
{\it interesting} than {\it right} in science. [A half-truth, but
  worth keeping in mind.]''  Studies at RSAA, ANU, of the Galaxy's
metal-poor populations by J.E.N. and D.Y. have been/are supported by
Australian Research Council DP0663562, DP0984924, DP120100475,
DP150100862, and FT140100554.  Parts of this research were conducted
by the Australian Research Council Centre of Excellence for All Sky
Astrophysics in 3 Dimensions (ASTRO 3D), through project number
CE170100013.

\newpage

%\bibliography{ump5,3d}
%\bibliographystyle{apj}

\clearpage
% Figure 1

\begin{figure}[htbp!]
\begin{center}
\includegraphics[width=8cm,angle=0]{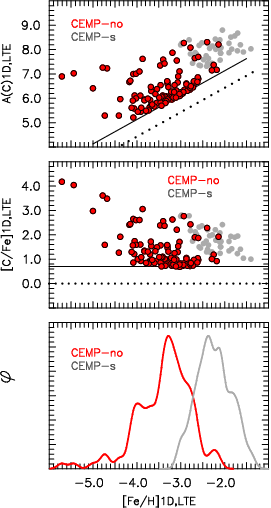}

 \caption{\label{fig:fig1}\small The dependence of $A$(C)$_{\rm
     1D,LTE}$, [C/Fe]$_{\rm 1D,LTE}$, and the generalized histogram of
   carbon abundance (adopting a Gaussion kernel of 0.10 dex), as a
   function of [Fe/H]$_{\rm 1D,LTE}$, for CEMP-no and CEMP-s stars
   (red and grey, respectively), based on the data of \citet{yoon16}.
   In the upper two panels the full line correspond to [C/Fe] = +0.7,
   the \citet{aoki07} boundary between CEMP and C-normal stars, while
   the dotted line is the [C/Fe] = 0.0 locus.}

\end{center}
\end{figure}

\clearpage
% Figure 2 

\begin{figure}[htbp!]
\begin{center}
\includegraphics[width=8cm,angle=0]{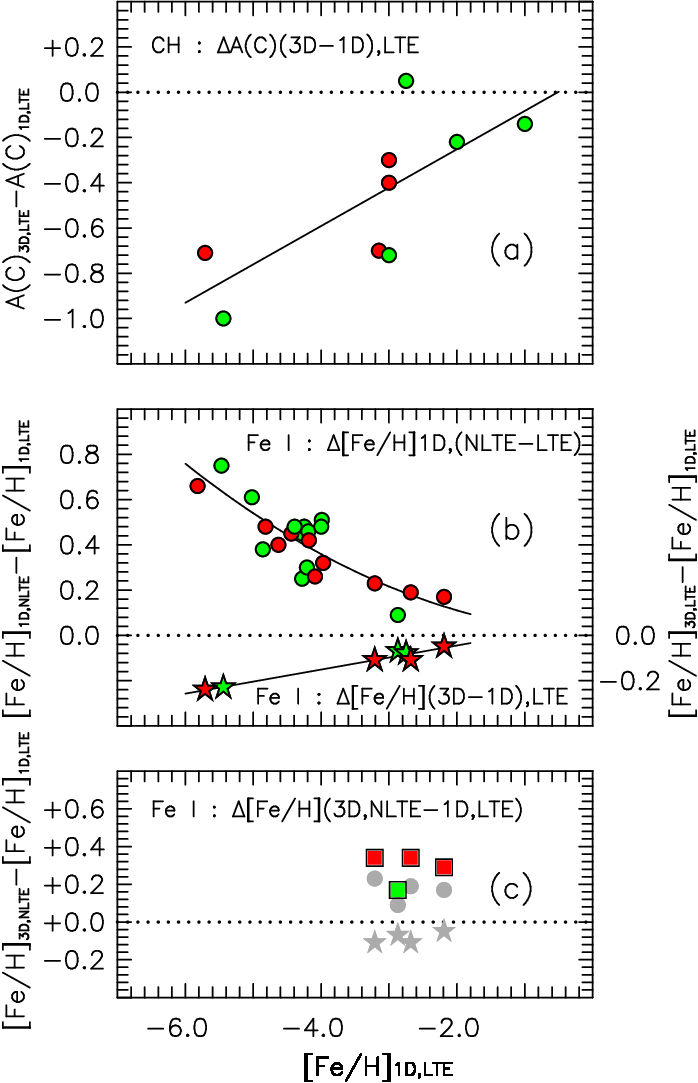}

  \caption{\label{fig:fig2}\small 3D and NLTE corrections as a
    function of [Fe/H]$_{\rm 1D,LTE}$.  In all panels, red and green
    symbols represent dwarfs and giants, respectively, from the work
    of \citet{amarsi16}, \citet{collet06, collet07, collet18},
    \citet{ezzeddine17}, \citet{frebel08}, \citet{gallagher16} and
    \citet{spite13} (see Table~\ref{tab:tab2}).  The panel (a)
    ordinate shows $\Delta$$A$(C)$_{\rm(3D-1D),LTE}$, while (b)
    presents $\Delta$[Fe/H]$_{\rm 1D,(NLTE-LTE)}$ and
    $\Delta$[Fe/H]$_{\rm(3D-1D),LTE}$. (The lines represent the
    least-squares best fits: (a) $\Delta$$A$(C)$_{\rm(3D-1D),LTE}$ =
    $A$(C)$_{\rm 3D,LTE}$ -- $A$(C)$_{\rm 1D,LTE}$ = 0.087 +
    0.170~[Fe/H]$_{\rm 1D,LTE}$, and (b) $\Delta$[Fe/H]$_{\rm
      1D,(NLTE-LTE)}$ = [Fe/H]$_{\rm 1D,NLTE}$ -- [Fe/H]$_{\rm
      1D,LTE}$ = 0.013 -- 0.011~[Fe/H]$_{\rm 1D,LTE}$ +
    0.019~[Fe/H]$_{\rm 1D,LTE}$$^{2}$ and
    $\Delta$[Fe/H]$_{\rm(3D-1D),LTE}$ = [Fe/H]$_{\rm 3D,LTE}$ --
    [Fe/H]$_{\rm 1D,LTE}$ = 0.061 + 0.053~[Fe/H]$_{\rm 1D,LTE}$).  In
    panel (c) the grey symbols repeat the \citet{amarsi16} [Fe/H]
    (3D--1D),LTE and 1D,(NLTE-LTE) corrections from panel (b), while
    the squares stand for $\Delta$[Fe/H] = [Fe/H]$_{\rm 3D,NLTE}$ --
    [Fe/H]$_{\rm 1D,LTE}$. We adopt [Fe/H]$_{\rm 3D,NLTE}$ --
    [Fe/H]$_{\rm 1D,NLTE}$ = 0.12.  See text for discussion.}
\end{center}
\end{figure}

\clearpage
% Figure 3 

\begin{figure}[htbp!]
\begin{center}
\includegraphics[width=6cm,angle=0]{f3.eps}

  \caption{\label{fig:fig3}\small The comparison between the 3D and 1D
    carbon abundances determined from analysis of the CH G-band and
    C~I high-excitation lines: (a) $A$(CI)$_{\rm 1D,NLTE}$ vs
    $A$(CH)$_{\rm 1D,LTE}$, (b) $A$(CI)$_{\rm 3D,LTE}$
    vs. $A$(CH)$_{\rm 3D,LTE}$, and (c) $\Delta$$A$(C) = $A$(CI)$_{\rm
      3D,NLTE}$ -- $A$(CH)$_{\rm 3D,LTE}$ as a function of
    [Fe/H]$_{\rm 1D,LTE}$. The red and green symbols represent stars
    for which CH-based abundances come from the present work and the
    literature, respectively.  In (a) and (b) the full line represents
    the 1-1 relation, while in (c) the wide red line is the linear
    least-squares best fit to the data.  See text for discussion.}
    
\end{center}
\end{figure}

\clearpage 
% Figure 4 

\begin{figure}[htbp!]
\begin{center}
\includegraphics[width=15cm,angle=0]{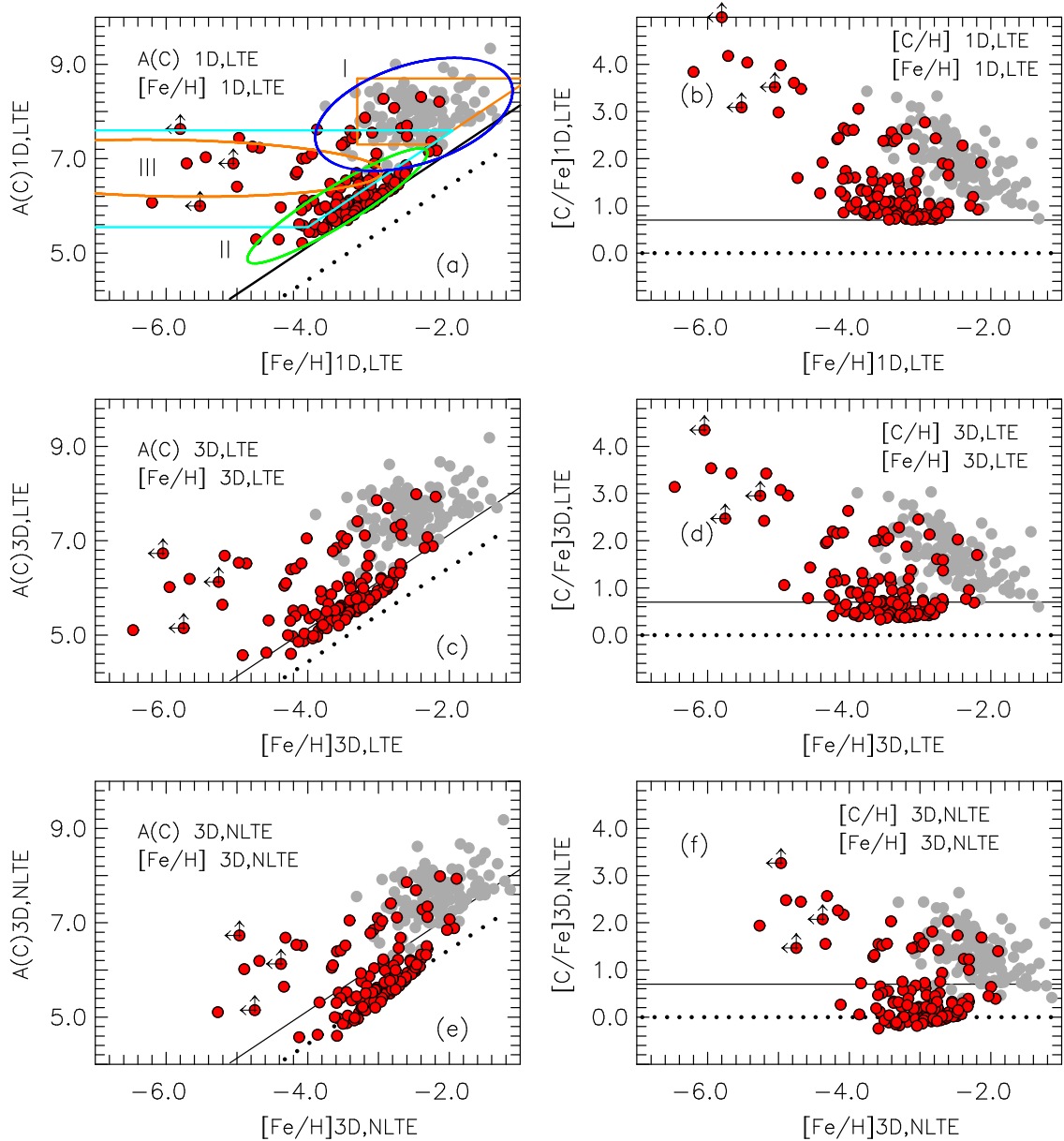}

  \caption{\label{fig:fig4}\small Carbon vs. iron abundances for
    CEMP-no (red symbols) and CEMP-s (grey symbols) stars as a
    function of 1D, 3D, LTE, and NLTE, to demonstrate how the
    distributions of C and Fe abundances depend on the underlying
    assumptions of the model atmosphere analysis.  The color for each
    individual star is as determined in the 1D,LTE uppermost panels (a
    and b). Data are taken from \citet{yoon16} and our
    Table~\ref{tab:tab4}.  In all panels the thin full and dotted
    horizontal and inclined lines refer to [C/Fe] = +0.7 and 0.0 loci.
    The blue, green, and orange ellipses in the top-left panel
    delineate the Yoon et al. Groups I, II, and III, respectively,
    while the horizontal orange and light blue lines (truncated by
    {C/Fe] = +1.0 loci) represent the ``high carbon band'' (orange)
    and ``low carbon band'' (light blue) of \citet{spite13},
    \citet{bonifacio15, bonifacio18}, and \citet{caffau18}.}}
\end{center}
\end{figure}

\clearpage
% Figure 5 
\begin{figure}[htbp!]
\begin{center}
\includegraphics[width=15cm,angle=0]{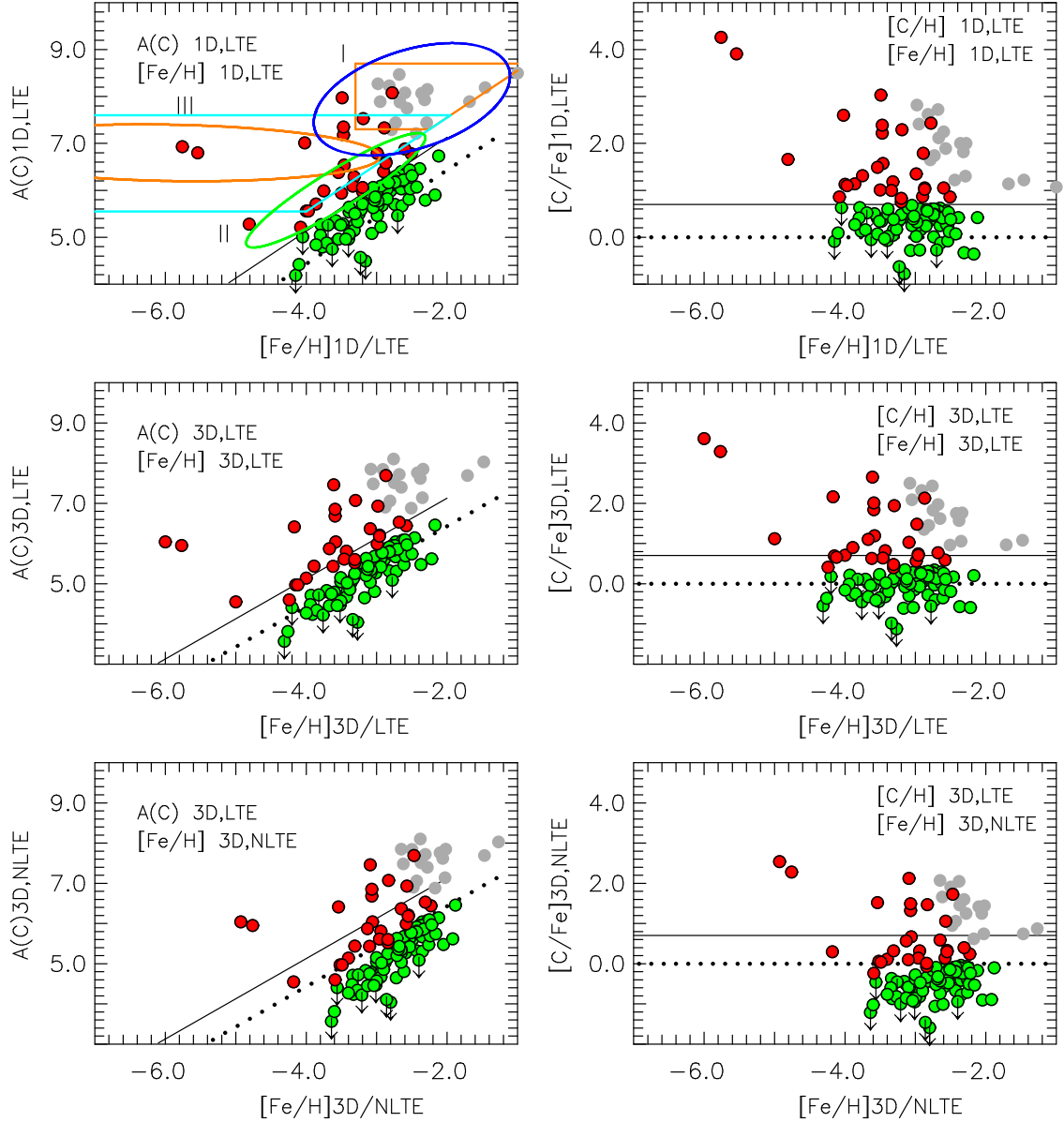}

  \caption{\label{fig:fig5}\small Carbon vs. iron abundances for the
    data sample of \citet{yong13a}, where the format is the same as
    that of Figure~\ref{fig:fig4}, with the exception that the green
    symbols represent 1D,LTE C-normal stars.  The color for each
    individual star is as determined in the 1D,LTE uppermost panels.
    The thin full and dotted lines refer to the [C/Fe] = +0.7 and 0.0 loci.
    See text for discussion.}

\end{center}
\end{figure}

\clearpage
% Figure 6

\begin{figure}[htbp!]
\begin{center}
\includegraphics[width=12.0cm,angle=0]{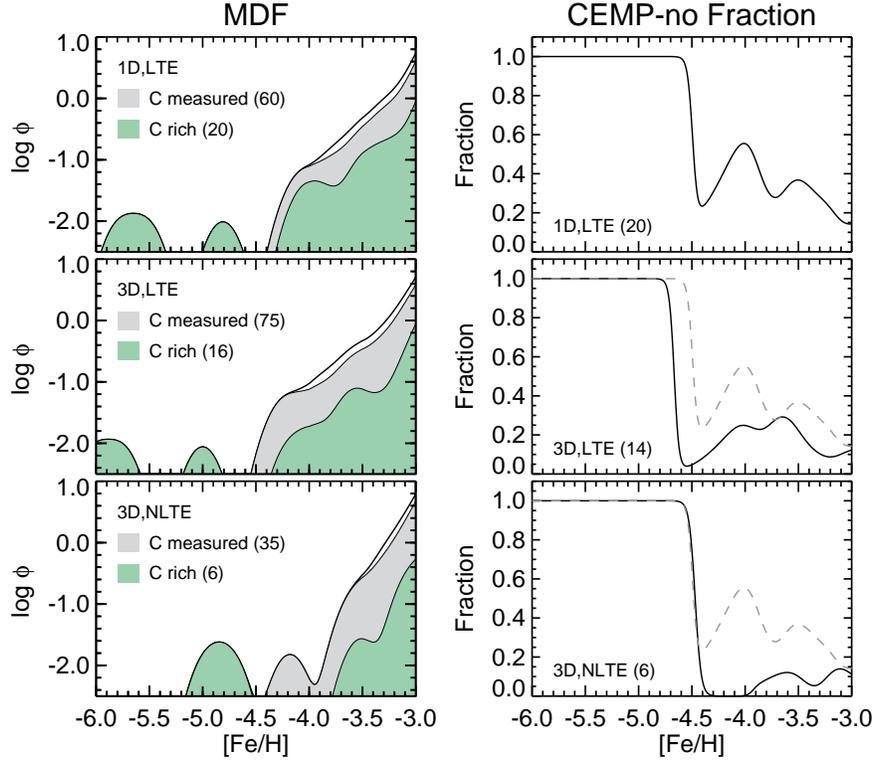}

  \caption{\label{fig:fig6}\small Metallicity distribution function
    (left) and CEMP-no fraction (right) (Gaussian histograms with
    kernel 0.3 dex) as a function of [Fe/H]$_{\rm 1D,LTE}$ (top),
    [Fe/H]$_{\rm 3D,LTE}$ (middle), and [Fe/H]$_{\rm 3D,NLTE}$
    (bottom).  See text for details.}

\end{center}
\end{figure}

\clearpage
% Figure 7 

\begin{figure}[htbp!]
\begin{center}
\includegraphics[width=13cm,angle=0]{f7.eps}

  \caption{\label{fig:fig7} \small Generalized histograms for (1D,LTE)
    [Na/Fe], [Mg/Fe], [Al/Fe], and [Ca/Fe] abundances (with Gaussian
    kernels of 0.30 dex).  The red thicker lines are for the CEMP-no
    stars of \citet{yoon16} and Table 4 of the present work, while the
    thinner black lines represent the C-normal stars of
    \citet{yong13a}. The legends contain the Group memberships of the
    samples, where the details in the leftmost panels apply to all of
    the panels in that row.  In the legends in the bottom two rows,
    [C/Fe] is the 1D,LTE value.  The number in each panel indicates
    the number of stars included in the histogram.}

\end{center}
\end{figure}

\clearpage
% Figure 8 

\begin{figure}[htbp!]
\begin{center}
\includegraphics[width=8cm,angle=0]{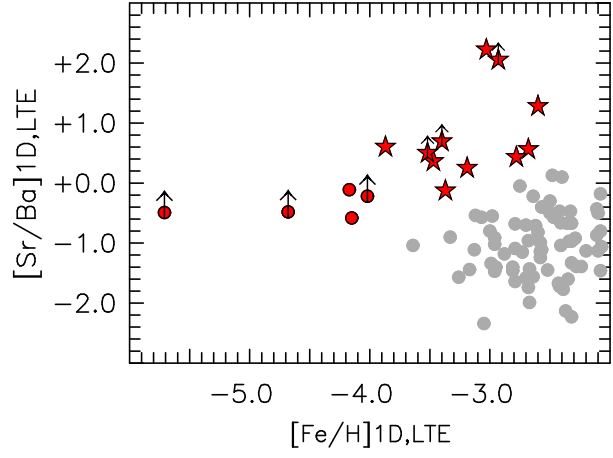}

  \caption{\label{fig:fig8}\small [Sr/Ba]$_{\rm 1D,LTE}$
    vs. [Fe/H]$_{\rm 1D,LTE}$.  Red star symbols and circles represent
    Group I, CEMP-no and Group III, CEMP-no stars, respectively, while
    grey circles stand for Group I CEMP-s stars.}

\end{center}
\end{figure}

\clearpage
% Figure 9
 
\begin{figure}[htbp!]
\begin{center}
\includegraphics[width=13cm,angle=0]{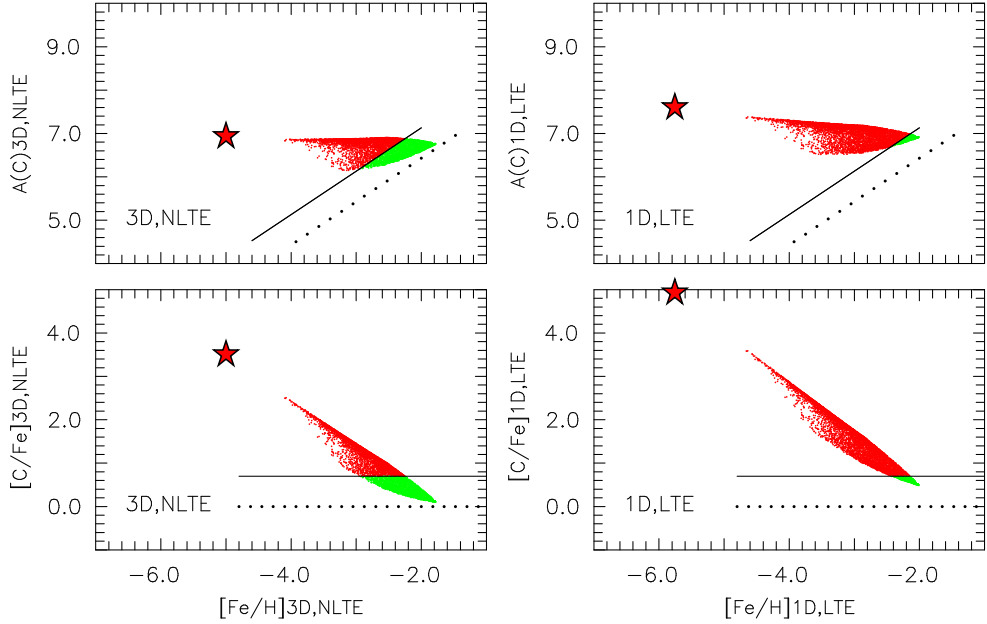}

  \caption{\label{fig:fig9}\small Toy model simulations assuming
    3D,NLTE (left) and 1D,LTE (right) adopting a C-rich population
    with model [Fe/H] = --5.0 and [C/Fe] = +3.5 (represented by the
    red star symbol in each panel).  The upper and lower panels
    present $A$(C) and [C/Fe] vs. [Fe/H], respectively, for the
    composite model population.  The full and dotted lines refer to the [C/Fe] =
    +0.7 and 0.0 loci, respectively, while green and red symbols represent
    stars with [C/Fe] below and above +0.7 dex.}

\end{center}
\end{figure}

\clearpage
% Figure 10 

\begin{figure}[htbp!]
\begin{center}
\includegraphics[width=13cm,angle=0]{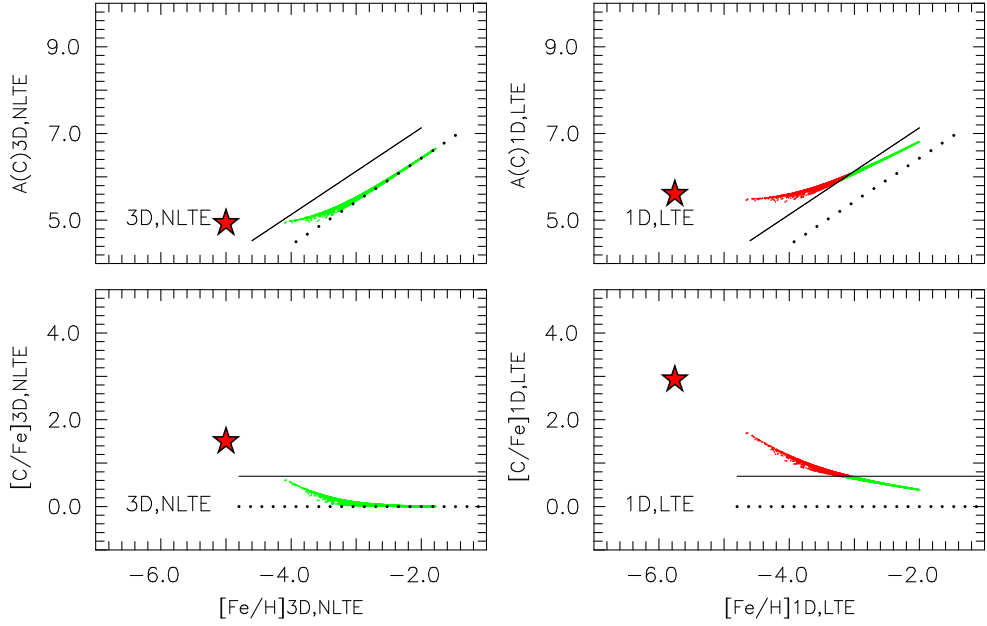}

  \caption{\label{fig:fig10}\small Toy model simulations assuming
    3D,NLTE (left) and 1D,LTE (right) adopting a C-rich population
    with model [Fe/H] = --5.0 and [C/Fe] = +1.5. The format is the
    same as in Figure~\ref{fig:fig9}, while the model assumes a
    considerably lower value of [C/Fe] for the C-rich population than
    in Figure~\ref{fig:fig9}.}

\end{center}
\end{figure}

\clearpage
% Figure 11 

\begin{figure}[htbp!]
\begin{center}
\includegraphics[width=10cm,angle=0]{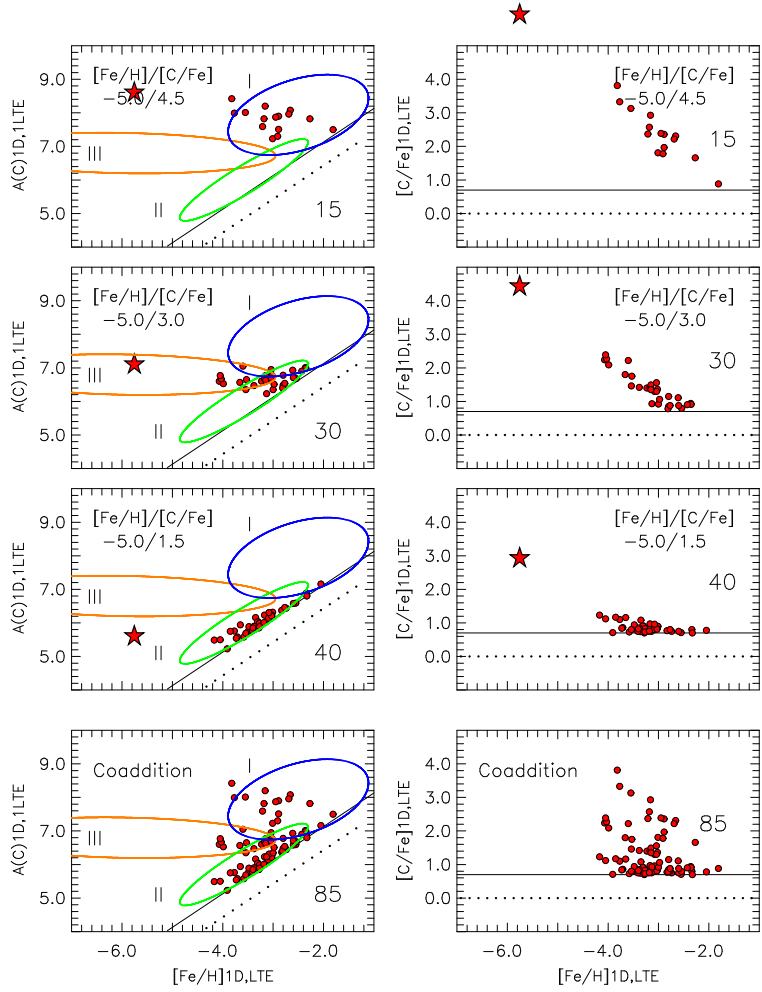}

  \caption{\label{fig:fig11}\small Toy model simulations for
    $A$(C)$_{\rm 1D,LTE}$ (left) and [C/Fe]$_{\rm 1D,LTE}$ (right)
    vs. [Fe/H]$_{\rm 1D,LTE}$, respectively.  The upper three rows
    present results obtained with the adopted [Fe/H]/[C/Fe] pairs for
    the C-rich population shown in each row, while the bottom row
    contains the superset of the above upper three panels.  The
    isolated number indicates the number of stars plotted in each
    panel.}

\end{center}
\end{figure}

\clearpage
% Table 1

\begin{deluxetable}{ll}                                         
\tablecolumns{2}                                                                    
\tablewidth{0pt}
\rotate
\tabletypesize{\footnotesize}
\tablecaption{\label{tab:tab1}  Major Milestones in the Study of CEMP-no Stars}              \tablehead{                                                                        
\colhead  {Milestone} &                         {Authors\tablenotemark{a}} \\  
  } 

\startdata 

Discovery of very metal-poor stars (\feh $<$ --2.0) with anomalously strong CH $\lambda$4300~{\AA} features & 1 \\ 
High dispersion abundance analyses reveal distinct C-rich subclasses & 2,3,4,5,6 \\
Taxonomy: [C/Fe]$_{\rm CEMP}$ $>$ +0.7; and subclasses CEMP-r, CEMP-s, CEMP-r/s, and CEMP-no & 7,8,9 \\
Taxonomy: Many CEMP-no stars have large supersolar abundances of N, O, Na, Mg, and Al & 10,11 \\ 
relative to Fe, but not of the heavy-neutron-capture elements (in particular, [Ba/Fe] $<$ 0.) &    \\ 
Taxonomy: Two distinct peaks in the [Fe/H] and $A$(C) histograms, populated principally by CEMP-no and CEMP-s stars  & 10,12,13,14,15 \\
Taxonomy: Essentially all CEMP stars with \feh \ltsima --3.3 belong to the CEMP-no subclass & 10,12 \\ 
Taxonomy: Two subgroups of CEMP-no stars exist in $A$(C)--[Fe/H] space & 16 \\ 
For CEMP-no stars, [C/Fe] increases strongly as [Fe/H] decreases & 17 \\
Discovery of C-rich stars having \feh $<$ --5.5 (assumed to be CEMP-no stars)  & 18,19,20 \\  
14 halo stars currently known to have \feh $<$ --4.5.  At least 11 of them are C-rich & 11,21,22,23,24,25,26 \\
CEMP-no main-sequence-turnoff stars with [Fe/H] $<$ --4.0 have A(Li) $<$ 2.0 & 14,27 \\ 
The earliest two observed stellar populations formed by cooling of C-rich and C-normal clouds & 10,28,29,30 \\ 
Suggested origin of CEMP-no enrichment: mixing and fallback stellar models (in minihalos) & 31,32,33,(34) \\       
Suggested origin of CEMP-no enrichment: spinstars & 35,36,37 \\
Suggested origin of CEMP-no enrichment: mixing and fallback + spinstars & 38 \\
Suggested origin of CEMP-no enrichment: binarity & 39 \\
Radial velocity monitoring supports CEMP-no binary fraction similar to that of C-normal halo stars & 10,40,41 \\ 
Most recent radial velocity monitoring reports that CEMP-no binary fraction is larger than C-normal halo star value & 42 \\ 
The fraction of CEMP-no stars increases as {\feh} decreases, and as Galactocentric distance increases  & 43,44,45 \\ 
The ratio of CEMP-no to CEMP-s stars reported to increase with increasing Galactocentric distance & 46,47,48 \\ 
CEMP-no stars exist in the Milky Way's dwarf ultra-faint galaxy satellites Bootes I and Segue 1  & 49,50 \\ 
Discovery of Damped Lyman-$\alpha$ Systems with enhanced [C/Fe] in quasar Ly$\alpha$ forests & 51,52,53 \\
Discussion of CEMP-no stars within the framework of the formation of the first galaxies & 54,55,56 \\
\enddata
\tablenotetext{a} {Authors: 
1 = \citet{beers92}, 
2 = \citet{sneden94}, 
3 = \citet{mcwilliam95}, 
4 = \citet{barbuy97}, 
5 = \citet{norris97a,norris97b}, 
6 = \citet{bonifacio98}, 
7 = \citet{aoki02,aoki07}, 
8 = \citet{beers05}, 
9 = \citet{ryan05}, 
10 = \citet{norris13}, 
11 = \citet{fn15}, 
12 = \citet{aoki10}, 
13 = \citet{spite13},
14 = \citet{bonifacio18}, 
15 = \citet{caffau18}
16 = \citet{yoon16}, 
17 = \citet{rossi99}, 
18 = \citet{christlieb02},
19 = \citet{frebel05},
20 = \citet{keller14}, 
21 = \citet{frebel15}, 
22 = \citet{caffau16},
23 = \citet{aguado_2},
24 = \citet{aguado_1},
25 = \citet{starkenburg18},
26 = \citet{nordlander19}, 
27 = \citet{frebel08},
28 = \citet{frebel07},
29 = \citet{schneider12},
30 = \citet{chiaki17},
31 = \citet{umeda03}, 
32 = \citet{iwamoto05}, 
33 = \citet{nomoto13}, 
34 = \citet{cooke14},
35 = \citet{meynet06}, 
36 = \citet{maeder15a},
37 = \citet{maeder15b},
38 = \citet{takahashi14},
39 = \citet{suda04}, 
40 = \citet{starkenburg14}, 
41 = \citet{hansent16}, 
42 = \citet{arentsen18},
43 = \citet{frebel06},
44 = \citet{carollo12},
45 = \citet{lee13},
46 = \citet{carollo14},
47 = \citet{lee17},
48 = \citet{hansen19},
49 = \citet{gilmore13}, 
50 = \citet{norris10}, 
51 = \citet{cooke11},
52 = \citet{cooke12},
53 = \citet{carswell12},
54 = \citet{becker12},
55 = \citet{sarmento17},
56 = \citet{sharma18}}
\end{deluxetable}

\clearpage
% Table 2
\vspace{-15mm}
\begin{deluxetable}{lccrrrrrrc}
\tablecolumns{10}        
\tabletypesize{\scriptsize}
\tablecaption{\label{tab:tab2} Literature [Fe/H] and $A$(C) values}                  
\tablehead{\colhead   {Star}    & {\teff} &  {\logg}    & [Fe/H]  & [Fe/H]   & [Fe/H]  & [Fe/H]  & $A$(C)   & $A$(C)      & Source\tablenotemark{a} \\
                     & {(K)}   &             & $_{\rm 1D,LTE}$  & $_{\rm 1D,NLTE}$  & $_{\rm 3D,LTE}$  & $_{\rm 3D,NLTE}$ & $_{\rm 1D,LTE}$ & $_{\rm 3D,LTE}$   & \\   
       {(1)}         & {(2)}   & {(3)}       & {(4)}   & {(5)}    & {(6)}   & {(7)}   & {(8)}  & {(9)}   & {(10)} 
}
\startdata 
HD~84937               & 6356   &     4.1     &   $-$2.19 &    $-$2.02 &   $-$2.24 & $-$1.90 &   ... &     ... & 1   \\ 
HD~122563              & 4587   &     1.6     &   $-$2.87 &    $-$2.78 &   $-$2.94 & $-$2.70 &   ... &     ... & 1   \\ 
HD~122563              & 4600   &     1.6     &   $-$2.75 &      ... &   $-$2.83 &   ... &  5.28 &    5.33 & 2   \\ 
HD~140283              & 5591   &     3.6     &   $-$2.68 &    $-$2.49 &   $-$2.79 & $-$2.34 &   ... &     ... & 1   \\  
G~64-12                & 6435   &     4.3     &   $-$3.21 &    $-$2.98 &   $-$3.32 & $-$2.87 &   ... &     ... & 1   \\ 
CD~$-$38$^{\circ}\,245$ & 4700   &     2.0     &   $-$4.28 &    $-$4.03 &     ... &  ...  &   ... &     ... & 3   \\ 
CS~22949-037           & 4800   &     1.9     &   $-$3.99 &    $-$3.48 &     ... &  ...  &   ... &     ... & 3   \\ 
CS~30336-049           & 4685   &     1.4     &   $-$4.21 &    $-$3.91 &     ... &  ...  &   ... &     ... & 3   \\ 
HE~0057$-$5959         & 5200   &     2.8     &   $-$4.28 &    $-$3.83 &     ... &  ...  &   ... &     ... & 3   \\   
HE~0107$-$5240         & 5130   &     2.2     &   $-$5.44 &      ... &   $-$5.67 &  ...  &  6.81 &    5.81 & 4   \\ 
HE~0107$-$5240         & 5050   &     2.3     &   $-$5.47 &    $-$4.72 &     ... &  ...  &   ... &     ... & 3   \\  
HE~0233$-$0343         & 6020   &     3.4     &   $-$4.44 &    $-$3.99 &     ... &  ...  &   ... &     ... & 3   \\   
HE~0557$-$4840         & 4800   &     2.4     &   $-$4.86 &    $-$4.48 &     ... &  ...  &   ... &     ... & 3   \\   
HE~1310$-$0536         & 5000   &     1.9     &   $-$4.25 &    $-$3.77 &     ... &  ...  &   ... &     ... & 3   \\   
HE~1327$-$2326         & 6190   &     3.9     &   $-$5.71 &      ... &   $-$5.95 &  ...  &  6.84 &    6.13 & 4,5 \\
HE~1327$-$2326         & 6130   &     3.7     &   $-$5.82 &    $-$5.16 &     ... &  ...  &   ... &     ... & 3   \\ 
HE~1424$-$0241         & 5140   &     2.8     &   $-$4.19 &    $-$3.73 &     ... &  ...  &   ... &     ... & 3   \\  
HE~2139$-$5432         & 5270   &     3.2     &   $-$4.00 &    $-$3.52 &     ... &  ...  &   ... &     ... & 3   \\  
HE~2239$-$5019         & 6000   &     3.5     &   $-$4.18 &    $-$3.76 &     ... &  ...  &   ... &     ... & 3   \\  
SD~0140+2344\tablenotemark{b}           & 5600   &     4.6     &   $-$4.09 &    $-$3.83 &     ... &  ...  &   ... &     ... & 3   \\  
SD~1029+1729\tablenotemark{b}           & 5811   &     4.0     &   $-$4.63 &    $-$4.23 &     ... &  ...  &   ... &     ... & 3   \\  
SD~1143+2020\tablenotemark{b}           & 6240   &     4.0     &   $-$3.15 &      ... &     ... &  ...  &  8.10 &    7.40 & 6  \\  
SD~1204+1201\tablenotemark{b}           & 5350   &     3.3     &   $-$4.39 &    $-$3.91 &     ... &  ...  &   ... &     ... & 3   \\  
SD~1313$-$0019\tablenotemark{b}         & 5100   &     2.7     &   $-$5.02 &    $-$4.41 &     ... &  ...  &   ... &     ... & 3   \\  
SD~1742+2531\tablenotemark{b}           & 6345   &     4.0     &   $-$4.82 &    $-$4.34 &     ... &  ...  &   ... &     ... & 3   \\  
SD~2209$-$0028\tablenotemark{b}         & 6440   &     4.0     &   $-$3.97 &    $-$3.65 &     ... &  ...  &   ... &     ... & 3   \\  
5131/2.2/$-$1.0       & 5131   &     2.2     &   $-$1.00 &      ... &     ... &  ...  &  7.52 &    7.40                  & 7   \\  
5035/2.2/$-$2.0       & 5035   &     2.2     &   $-$2.00 &      ... &     ... &  ...  &  6.52 &    6.30                  & 7   \\  
5128/2.2/$-$3.0       & 5128   &     2.2     &   $-$3.00 &      ... &     ... &  ...  &  5.52 &    4.80                  & 7   \\  
C-normal dwarfs       & 5900--6500 & 4.0--4.5 &  $-$3.00 &      ... &     ... &  ...  &  5.95 &    5.55                  & 8   \\  
CEMP-no dwarfs        & 5900--6500 & 4.0--4.5 &  $-$3.00 &      ... &     ... &  ...  &  6.80 &    6.50\tablenotemark{c} & 8   
\enddata
\tablenotetext{a} {Source: 1 = \citet{amarsi16}, 2 = \citet{collet18}, 3 = \citet{ezzeddine17}, 4 = \citet{collet06}, 5 = \citet{frebel08}, 6 = \citet{spite13}, 7 = \citet{collet07}, 8 = \citet[Section 4.2]{gallagher16}.} 
\tablenotetext{b} {SD~0140+2344 = SDSS~J0140+2344, SD~1029+1729 = SDSS~J1029+1729, SD~1143+2020 = SDSS~J1143+2020, SD~1204+1201 = SDSS~J1204+1201,
SD~1313$-$0019 = SDSS~J1313$-$0019, SD~1742+2531 = SDSS~J1742+2531, SD~2209$-$0028 = SDSS~J2209$-$0028}
\tablenotetext{c}  {We adopt the \citet{gallagher16} result for low $A$(C) $\sim$~6.8, which is more pertinent to CEMP-no stars.}
\end{deluxetable}

\clearpage
% Table 3

\begin{deluxetable}{lcccccrrrrr}
\tablecolumns{11}        
\tablewidth{0pt}
\tabletypesize{\footnotesize}
\tablecaption{\label{tab:tab3}  Carbon abundances\tablenotemark{a} from C~I lines and the CH G-band}                  
\tablehead{\colhead   
       {Star}  & {\teff} & {\logg} & [Fe/H]          & $A$(C~I)          & $A$(C~I)          & $A$(CH)           & $A$(CH)           & $A$(C~I)$_{\rm 1D,NLTE}$  & $S/N$   & Source\tablenotemark{b} \\
               & {(K)}   &         & $_{\rm 1D,LTE}$  & $_{\rm 1D,LTE}$  & $_{\rm 1D,NLTE}$  & $_{\rm 1D,LTE}$  & $_{\rm 3D,LTE}$  & $-$$A$(CH)$_{\rm 3D,LTE}$  &         &        \\   
       {(1)}   & {(2)}   & {(3)}   & {(4)}           & {(5)}           & {(6)}           & {(7)}           & {(8)}           & {(9)}                  & {(10)}  & {(11)} \\ }
\startdata 
HD~84937      &  6357 &   4.07 &  $-$2.11 &   6.57 &   6.32 &   6.56 &     6.29 &      0.03 & 500  &  1 \\
HD~140283     &  5849 &   3.72 &  $-$2.38 &   6.31 &   6.07 &   6.21 &     5.89 &      0.18 & 470  &  2 \\
HD~188031     &  6234 &   4.16 &  $-$1.72 &   6.88 &   6.67 &   6.91 &     6.70 &   $-$0.03 & 180  &  3 \\
HD~215801     &  6071 &   3.83 &  $-$2.28 &   6.36 &   6.10 &   6.46 &     6.16 &   $-$0.06 & 180  &  3 \\
BD~$-$13 3442   &  6366 &   3.99 &  $-$2.69 &   6.18 &   5.85 &$<$6.16 & $<$ 5.79 & $>$~0.06 & 200  &  3 \\
CD~$-$24 17504  &  6338 &   4.32 &  $-$3.21 &   5.85 &   5.55 &   6.06 &     5.60 &   $-$0.05 & 280  &  4 \\
CD~$-$24 17504  &  6228 &   3.90 &  $-$3.41 &   5.71 &   5.45 &   6.12 &     5.63 &   $-$0.18 & 510  &  5 \\
CD~$-$35 14849  &  6294 &   4.26 &  $-$2.34 &   6.42 &   6.17 &   6.51 &     6.20 &   $-$0.03 & 180  &  3 \\
CD~$-$42 14278  &  6085 &   4.39 &  $-$2.03 &   6.58 &   6.41 &   6.76 &     6.50 &   $-$0.09 &  90  &  6 \\
CD~$-$71 1234   &  6325 &   4.18 &  $-$2.38 &   6.32 &   6.07 &   6.31 &     5.99 &      0.08 & 310  &  4 \\
G~4-37        &  6308 &   4.25 &  $-$2.45 &   6.30 &   6.03 &   6.46 &     6.13 &   $-$0.10 & 150  &  4 \\
G~11-44       &  6178 &   4.35 &  $-$2.03 &   6.62 &   6.45 &   6.71 &     6.45 &      0.00 & 120  &  6 \\
G~13-9        &  6343 &   4.01 &  $-$2.29 &   6.52 &   6.22 &   6.46 &     6.16 &      0.06 & 280  &  7 \\
G~24-3        &  6084 &   4.23 &  $-$1.62 &   6.69 &   6.52 &   6.86 &     6.67 &   $-$0.15 & 240  &  8 \\
G~29-23       &  6194 &   4.04 &  $-$1.69 &   6.86 &   6.64 &   6.86 &     6.66 &   $-$0.02 & 200  &  8 \\
G~48-29       &  6489 &   4.25 &  $-$2.60 &   6.05 &   5.76 & $<$6.46 & $<$6.11 &   $>$$-$0.35 & 240  &  4 \\
G~48-29       &  6489 &   4.25 &  $-$2.60 &   6.05 &   5.76 & $<$6.31 & $<$5.96 &   $>$$-$0.20 & 220  &  9 \\
G~59-27       &  6272 &   4.23 &  $-$1.93 &   6.85 &   6.63 &   6.86 &     6.62 &      0.01 & 210  &  4 \\
G~64-12       &  6435 &   4.26 &  $-$3.24 &   5.71 &   5.30 &   6.16 &     5.70 &   $-$0.40 & 700  & 10 \\
G~64-12       &  6463 &   4.26 &  $-$3.29 &   5.71 &   5.30 &   6.21 &     5.74 &   $-$0.44 & 700  & 11 \\ 
G~64-37       &  6570 &   4.40 &  $-$3.11 &   5.75 &   5.34 &   6.36 &     5.92 &   $-$0.58 & 700  & 10 \\
G~64-37       &  6570 &   4.40 &  $-$3.11 &   5.75 &   5.34 &   6.44 &     6.00 &   $-$0.66 & 700  & 11 \\ 
G~126-52      &  6396 &   4.20 &  $-$2.21 &   6.46 &   6.21 &   6.46 &     6.17 &      0.04 & 270  &  4 \\
G~166-54      &  6407 &   4.28 &  $-$2.58 &   6.07 &   5.75 &   6.16 &     5.81 &   $-$0.06 & 240  &  4 \\
G~186-26      &  6417 &   4.42 &  $-$2.54 &   6.19 &   5.91 &   6.32 &     5.98 &   $-$0.07 & 110  &  3 \\
LP~635-14     &  6367 &   4.11 &  $-$2.39 &   6.43 &   6.14 &   6.46 &     6.14 &      0.00 & 250  &  4 \\
LP~651-4      &  6371 &   4.20 &  $-$2.63 &   6.12 &   5.81 &   6.21 &     5.85 &   $-$0.04 & 240  &  4 \\
\enddata
\tablenotetext{a} {Here and in Section~\ref{sec:ci}, $A$(C~I) and $A$(CH) refer to $A$(C) abundances determined from C~I lines and the CH G-band, respectively}
\tablenotetext{b} {Source: 1 = UVES, 266.D-5655(A), 2 = UVES, 165.N-0276(A), 3 = UVES, 95.D-0504(A), 4 = UVES, 73.D-0024(A), 5 = Results from \citet{jacobson15}, 6 = UVES, 86.D-0871(A), 7 = UVES, 67.D-0086(A), 8 = UVES, 71.B-0529(A), 9 = UVES,170.D-0010(G), 10 = HIRES,PI Melendez, 11 = Results from \citet{placco16}}
\end{deluxetable}

\clearpage 
% Table 4

\begin{deluxetable}{lcccccrr}                                                          
\tablecolumns{8}        
\tablewidth{0pt}
\tabletypesize{\small}
\tablecaption{\label{tab:tab4} {\teff}, {\logg}, and carbon and iron abundances for eight recently reported CEMP-no stars}                  
\tablehead{\colhead   {Star} & {\teff} & {\logg} & [Fe/H] & [C/Fe]\tablenotemark{a} & $A$(C)     & [Ba/Fe] & Source\tablenotemark{b} \\
                             & {(K)}   &         & 1D,LTE & 1D,LTE & 1D,LTE   &         &                         \\ 
                      {(1)}  & {(2)}   & {(3)}   & {(4)}  & {(5)}  & {(6)}    & {(7)}   & {(8)}                   \\           
}
\startdata 
G~64-12                                &  6463  & 4.26  &  $-$3.29 &    +1.07 &    6.21 & $-$0.06  &   1   \\
G~64-37                                &  6570  & 4.40  &  $-$3.11 &    +1.12 &    6.44 & $-$0.35  &   1   \\
SD~1341+4741\tablenotemark{c}          &  5450  & 2.50  &  $-$3.20 &    +1.00 &    6.23 & $-$0.73  &   2   \\ 
Bootes I-119                           &  4770  & 1.40  &  $-$3.33 &    +2.42 &    7.52 & $-$1.00  &   3,4 \\ 
Pisces II-10694                        &  4130  & 0.80  &  $-$2.60 &    +1.58 &    7.48 & $-$1.10  &   5   \\ 
Segue 1-7                              &  4960  & 1.90  &  $-$3.52 &    +2.38 &    7.29 & $<-$0.96 &   6   \\ 
Segue 1 SD~1006+1602\tablenotemark{c}  &  5484  & 3.30  &  $-$3.60 &    +1.20 &    6.03 & $<-$1.87 &   7   \\ 
Segue 1 SD~1006+1600\tablenotemark{c}  &  5170  & 2.50  &  $-$3.78 &    +0.91 &    5.54 & $<-$2.25 &   7   \\ 
\enddata
\tablenotetext{a}{Corrected for stellar evolutionary effects following \citet{placco14} (using http://www.nd.edu/~vplacco/carbon-cor.html)}
\tablenotetext{b}{Source: 
1 = \citet{placco16b}, 
2 = \citet{bandy18}, 
3 = \citet{gilmore13}, 
4 = \citet{lai11}, 
5 = \citet{spite18}, 
6 = \citet{norris10}, 
7 = \citet{frebel14}}
\tablenotetext{c}{SD~1341+4741 = J134144+474128, SD~1006+1602 = SDSS J100652+160235, SD~1006+1600 = SDSS J100639+160008}
\end{deluxetable}

\clearpage 
% Table 5

\begin{deluxetable}{lcccccrcr}                                                          
\tablecolumns{9} \tablewidth{0pt}
\tabletypesize{\footnotesize}
\tablecaption{\label{tab:tab5} Data for 19 CEMP-no, Group I Stars}                  
\tablehead{\colhead   {Star}  & {\teff} & {\logg} & [Fe/H] & [C/Fe] & $A$(C)  & [Ba/Fe] & Binary? & Source\tablenotemark{a}\\ 
                              & {(K)}   &         & 1D,LTE & 1D,LTE & 1D,LTE   &       &        &         \\ 
                      {(1)}   & {(2)}   & {(3)}   & {(4)}  & {(5)}  & {(6)}    & {(7)} & {(8)}  & {(9)}   \\           
}
\startdata 
CS~22943-201      & 5970 &   2.45 &  $-$2.68 &   1.90 &   7.65 &  $-$0.54 & ... &  1   \\ 
CS~22957-027      & 5220 &   2.65 &  $-$3.19 &   2.63 &   7.87 &  $-$0.81 & Yes &  2,3 \\ 
CS~22958-042      & 5760 &   3.55 &  $-$3.40 &   2.56 &   7.59 &  $-$0.61 & ... &  4,1\tablenotemark{b} \\ 
CS~29498-043      & 4440 &   0.50 &  $-$3.87 &   3.06 &   7.62 &  $-$0.49 & No  &  3   \\ 
HE~0100$-$1622    & 5400 &   3.00 &  $-$2.93 &   2.77 &   8.27 &  $-$1.80 & ... &  5   \\ 
HE~0219$-$1739    & 4238 &   0.47 &  $-$3.09 &   2.21 &   7.55 &  $-$1.39 & Yes &  3   \\ 
HE~0405$-$0526    & 5083 &   3.86 &  $-$2.18 &   0.92 &   7.17 &  $-$0.22 & No  &  3   \\ 
HE~1133$-$0555    & 5526 &   1.31 &  $-$2.40 &   2.28 &   8.31 &  $-$0.58 & No  &  3   \\ 
HE~1150$-$0428    & 5208 &   2.54 &  $-$3.47 &   2.39 &   7.35 &  $-$0.48 & Yes &  3   \\ 
HE~1302$-$0954    & 5120 &   2.40 &  $-$2.25 &   1.19 &   7.37 &  $-$0.53 & No  &  3   \\ 
HE~1330$-$0354    & 6257 &   4.13 &  $-$2.29 &   1.01 &   7.15 &  $-$0.52 & ... &  6   \\  
HE~1410+0213      & 5000 &   2.00 &  $-$2.14 &   1.92 &   8.21 &  $-$0.26 & No  &  3   \\ 
HE~1456+0230      & 5664 &   2.20 &  $-$3.37 &   2.37 &   7.43 &  $-$0.19 & ... &  7   \\ 
HE~2202$-$4831    & 5331 &   2.95 &  $-$2.78 &   2.43 &   8.08 &  $-$1.28 & ... &  8   \\ 
HE~2319$-$5228    & 4900 &   1.60 &  $-$2.60 &   1.87 &   7.70 &  $-$3.00 & ... &  9   \\ 
SDSS~J1422+0031   & 5200 &   2.20 &  $-$3.03 &   1.71 &   7.11 &  $-$1.18 & Yes & 10   \\
Bootes~I-119      & 4770 &   1.40 &  $-$3.33 &   2.42 &   7.52 &  $-$1.00 & ... & 11   \\ 
Pisces~II-10694   & 4130 &   0.80 &  $-$2.60 &   1.65 &   7.48 &  $-$1.10 & ... & 12    \\ 
Segue~1-7         & 4960 &   1.90 &  $-$3.52 &   2.38 &   7.29 &  $-$0.96 & ... & 13,14\tablenotemark{c} \\
\enddata
\tablenotetext{a} {1 = \citet{roederer14}, 2 = \citet{preston01}, 3 = \citet{hansent16}, 4 = \citet{sivarani06}, 5 = \citet{hansent15}, 6 = \citet{barklem05}, 7 = \citet{cohen13}, 8 = \citet{yong13a}, 9 = \citet{hansenc16}, 10 = \citet{arentsen18}, 11 = \citet{gilmore13}, 12 = \citet{spite18}, 13 = \citet{norris10}, 13 = \citet{starkenburg14}}
\tablenotetext{b} {Observations of \citet{sivarani06} and \citet{roederer14}, taken some 440 days apart, agree to within $\sim$~2~km s$^{-1}$}
\tablenotetext{c} {Observations of \citet{norris10} and \citet{starkenburg14}, taken some 1100 days apart, agree to within $\sim$~1~km s$^{-1}$}
\end{deluxetable}

\clearpage 
% Table 6

\begin{deluxetable}{lrrrrrrr  }                                                          
\tablecolumns{8} \tablewidth{0pt}
\tabletypesize{\footnotesize}
\tablecaption{\label{tab:tab6} The 14 Most Iron-poor Stars}
\tablehead{\colhead {Star}              &  {RA2000} & {Dec2000}     & {\teff} & {\logg}  &  {[Fe/H]}  & {[C/Fe]}   & {Sources\tablenotemark{a}} \\
                     {(1)}              &   {(2)}   & {  (3)}       & {(4)}   &  {(5)}   &   {(6) }   &   {(7)}    &  {(8)}   \\ 
}
\startdata 
SMSS~J0313$-$6708   & 03 13 00.4 & $-$67 08 39.3 & 5125    & 2.3    & $<$$-$7.30  &    $>$+4.9    &  1 \\ 
SMSS~J1605$-$1443\tablenotemark{b}   & 16 05 40.2 & $-$14 43 23.1 & 4850    & 2.0    &    $-$6.20  &       +3.9    &  2  \\ 
SDSS~J0815+4729\tablenotemark{b}     & 08 15 54.2 &   +47 29 47.8 & 6215    & 4.7    & $<$$-$5.80  &    $>$+5.0    &  3  \\ 
HE~1327$-$2326      & 13 30 06.0 & $-$23 41 49.7 & 6180    & 3.7    &    $-$5.66  &       +4.3    &  4  \\ 
SDSS~J0023+0307\tablenotemark{b}     & 00 23 14.0 &   +03 07 58.1 & 6224    & 4.8    & $<$$-$5.50  &    $>$+3.2    &  5  \\ 
HE~0107$-$5240      & 01 09 29.2 & $-$52 24 34.2 & 5100    & 2.2    &    $-$5.39  &       +3.7    &  6  \\ 
SDSS~J1035+0641     & 10 35 56.1 &   +06 41 44.0 & 6262    & 4.0    & $<$$-$5.07  &    $>$+3.5    &  7  \\ 
SDSS~J1313$-$0019   & 13 13 26.9 & $-$01 19 41.4 & 5200    & 2.6    &    $-$5.00  & $\sim$+3.0    &  8  \\ 
SDSS~J0929+0238\tablenotemark{b}     & 09 29 12.3 &   +02 38 17.0 & 5894    & 3.7    &    $-$4.97  &       +3.9    &  9  \\ 
SDSS~J1742+2531     & 17 42 59.7 &   +25 31 35.9 & 6345    & 4.0    &    $-$4.80  &       +3.6    & 10  \\ 
HE~0557$-$4840      & 05 58 39.3 & $-$48 39 56.8 & 4900    & 2.0    &    $-$4.75  &       +1.6    & 11  \\ 
SDSS~J1029+1729     & 10 29 15.2 &   +17 29 28.0 & 5811    & 4.0    &    $-$4.73  &    $<$+0.9    & 12  \\ 
HE~0233$-$0343      & 02 36 29.7 & $-$03 30 06.0 & 6100    & 3.4    &    $-$4.68  &       +3.5    & 13  \\ 
Pristine\_221.8781+9.7844 & 14 47 30.7 &   +09 47 03.7 & 5792 & 3.5 &    $-$4.66  &    $<$+1.8    &  14 \\ 
\enddata
\tablenotetext{a} {1 = \citet{keller14} 2 = \citet{nordlander19}, 3 = \citet{aguado_1}, 4 = \citet{frebel05}, \citet{aoki06}, 5 = \citet{aguado19}, \citet{frebel19} 6 = \citet{christlieb04}, 7 = \citet{bonifacio15}, 8 = \citet{frebel15}, 9 = \citet{caffau16}, 10 = \citet{bonifacio15}, 11 = \citet{norris07}, 12 = \citet{caffau12}, 13 = \citet{hansen14}, 14 = \citet{starkenburg18}}
\tablenotetext{b} {This star is included in the analysis in Section~\ref{sec:yoon}}
\end{deluxetable}
\end{document}